# A Scoping Review of ChatGPT Research in Accounting and Finance

July 2024

Mengming Michael Dong
College of Business, Missouri State University
michaeldong@missouristate.edu

Theophanis C. Stratopoulos
School of Accounting and Finance, University of Waterloo
tstratopoulos@uwaterloo.ca

Victor Xiaoqi Wang*
College of Business, California State University Long Beach
victor.wang@csulb.edu

ABSTRACT: This paper provides a review of recent publications and working papers on ChatGPT and related Large Language Models (LLMs) in accounting and finance. The aim is to understand the current state of research in these two areas and identify potential research opportunities for future inquiry. We identify three common themes from these earlier studies. The first theme focuses on applications of ChatGPT and LLMs in various fields of accounting and finance. The second theme utilizes ChatGPT and LLMs as a new research tool by leveraging their capabilities such as classification, summarization, and text generation. The third theme investigates implications of LLM adoption for accounting and finance professionals, as well as for various organizations and sectors. While these earlier studies provide valuable insights, they leave many important questions unanswered or partially addressed. We propose venues for further exploration and provide technical guidance for researchers seeking to employ ChatGPT and related LLMs as a tool for their research.

* Corresponding author

# A Scoping Review of ChatGPT Research in Accounting and Finance

## I. INTRODUCTION

In the summer of 2022, The Economist (2022) described Large Language Models (LLMs) - like OpenAI's GPT-3 - as uncanny[1] due to their capability to generate human language, which up to that point was considered the pinnacle of intelligence (Wolfram 2023). In less than a year, OpenAI introduced two major updates: GPT-3.5 on November 30, 2022, and GPT-4 on March 14, 2023.[2] Unlike previous technologies designed to automate routine and repetitive tasks, this new technology can potentially replace workers in highly educated, well-compensated white-collar occupations. The most advanced LLMs have exhibited characteristics of general-purpose technologies, suggesting they could bring about significant economic, social, and policy ramifications (Eloundou et al. 2023). These developments have sparked heated discussions within various industries, including accounting and finance, and an unprecedented adoption rate. Gartner predicts over 80% enterprise adoption by 2026, a sharp spike from 5% in 2023 (Cooney 2023). In comparison, enterprise systems and cloud computing took approximately eight and six years, respectively, to achieve a 15% adoption rate (Stratopoulos and Wang 2022).

The repercussions of ChatGPT and other LLMs are keenly felt in academic circles. In a relatively short period (from January 2022 to March 2024), 264 papers related to ChatGPT and other LLMs were uploaded to SSRN within the Accounting, Finance, or Economics networks. Given the surge in scholarly activity and the vital importance of this emerging technology, it becomes imperative to synthesize and analyze this burgeoning body of work. Our literature review serves multiple purposes. First, it seeks to capture the current state of the art in LLM-related research in accounting and finance. Synthesizing current studies provides practitioners and researchers valuable insights into the latest developments and applications. Second, it aims to identify gaps in the existing literature. By critically evaluating existing studies, this review pinpoints areas where further research opportunities are fruitful and abundant. Lastly, this review critically assesses the methodologies researchers employ using LLMs as research tools and offers guidance on appropriately and effectively leveraging these models while avoiding potential pitfalls.

Traditionally, literature reviews synthesize well-established and published bodies of knowledge. However, as highlighted earlier, the unprecedented pace of technological advancement necessitates a shift in our approach. Rather than solely reflecting on where the research has been, we must adopt a forward-looking perspective that aligns with the famous adage, "Skate to where the puck is going to be, not where it has been." In this context, our review covers published works and working papers on SSRN. While published works remain valuable, they often lag behind the cutting-edge developments reflected in working papers. By synthesizing insights from published works and working papers, our analysis captures the dynamic landscape of ChatGPT's role in accounting and finance. This comprehensive approach enables us to remain at the forefront of a rapidly evolving field and provide timely, relevant insights into the potential implications and applications of the groundbreaking technology.

---

[1] The term "uncanny valley" was introduced by Robotics professor Masahiro Mori in 1970 to capture the feelings of eeriness and revulsion in humans when confronted with humanlike machines (Wikipedia 2023).
[2] For more about ChatGPT and the models behind it, see Section II – Background.

This paper complements existing survey studies focusing more on the technical aspects of applying LLMs to related fields. For example, Li, Wang, Ding, and Chen (2023) provide an overview of existing LLMs for various finance tasks, as well as on how to finetune pre-trained LLMs or train domain-specific LLMs from scratch. They also offer guidance on key considerations while applying LLMs in finance, such as technical suitability, cost/benefit trade-offs, risks, and limitations. On the other hand, Hadi et al. (2023) discuss fundamental concepts of generative AI, the architecture of GPT, the history/evolution of LLMs, how to train LLMs and their applications. Min et al. (2023) survey recent advancements in pre-trained LLMs, focusing on their Natural Language Processing (NLP) capabilities. Siddik et al. (2023) provide a review of the applications of ChatGPT in Fintech. Ray (2023) reviews the background of ChatGPT and its general applications.

To enhance the coherence of this literature review, we use a framework inspired by recent reviews on the adoption of emerging technologies such as blockchain (Yang Li et al. 2018) and AI (Lee et al. 2023). At its core, our framework adopts an input-process-output model (Lee et al. 2023), where the *input* relates to motivation for adoption and focus area of application, the *process* to how the technology is used, and the *output* to implications of widespread adoption. More specifically, for the first component of our framework, we delve into the motivations driving the adoption of LLMs. Researchers are natural innovators and are not surprisingly among the earliest adopters who employ the new tool to exploit research opportunities that offer quick returns. Therefore, a systematic analysis of the foci of these studies would serve as a proxy for the areas (e.g., audit, financial reporting) where researchers identify as having the strongest motivations, reflected as initial, most accessible opportunities for applying LLMs. In the *process* phase, we survey how LLMs are employed in the context of accounting and finance. This involves an examination of the specific capabilities of LLMs that researchers leverage in their work (e.g., text generation, classification, and summarization) or how they should be used in accounting or finance practice as proposed by researchers. In the *output* phase, we organize studies into four groups based on their implied adoption maturity (the stage in the adoption cycle): conceptual papers, case studies, potential applications, and value realization. Additionally, we will examine the impact of LLMs on education and labor markets for this last phase because such impacts result from the widespread adoption of the technology.

Approaching this body of work through three interconnected perspectives enables a well-organized and unified categorization of studies. The first perspective reveals that researchers anticipate efficiency and effectiveness gains in nearly all accounting and finance domains. These studies are highly concentrated in four primary areas: audit, financial reporting, asset pricing and investment, and corporate finance. Early evidence suggests that professionals aided by LLMs will likely outperform their counterparts who do not use these advanced tools. This trend points towards a potential shift from conventional labor practices to workflows augmented by LLMs. While these studies underscore substantial benefits, they also emphasize the importance of cautiously proceeding and carefully considering the risks of adopting these emerging technologies.

A similar message emerges from the second, process-oriented perspective. LLMs often outperform traditional methods in tasks such as classification, sentiment analysis, and summarization. The greater efficiency suggests that researchers and professionals using LLMs would be more productive than their counterparts relying on older methods. This is consistent with evidence from educational studies and studies examining the implication of LLMs for the accounting and finance profession. These studies have shown that ChatGPT-4 can pass various professional exams (e.g., CPA, CMA), perform tasks at a level comparable to a human auditor, augment the abilities of financial analysts, and offer practical financial advice. Finally, the output-oriented perspective of our review indicates a shift in focus from the conceptual aspects of LLMs to their potential applications. This trend demonstrates growing confidence in LLM capabilities and a transition to mainstream adoption, as well as hints at a potential transformation in task execution across various domains.

In summary, the expanding research on ChatGPT and related LLMs within accounting and finance mirrors the growing enterprise adoption of these technologies. This burgeoning area of inquiry is abundant with unexplored questions, offering a fertile ground for scholarly investigation. Late in the paper, we propose numerous research avenues, which we believe could yield significant contributions to the theoretical understanding of technology adoption for LLMs as well as their practical applications to and implications for various fields of accounting and finance.

The rest of the paper is organized as follows. Section II provides background information on LLMs and ChatGPT. Section III describes the scope of our review and the methodology used. Section IV presents a descriptive analysis of the papers and synthesizes them from the lens of input, process, and output. Section V discusses these papers and their broader implications across various streams of literature, proposing venues for future research. Section VI concludes with closing remarks. The Appendix offers technical guidance on leveraging ChatGPT and LLMs as research tools.

## II. BACKGROUND

**2.1 LLMs**

An LLM is a machine learning model trained to understand, generate, and interact with human language.[3] The label "large" comes from the fact that such models have an enormous number of parameters, often on the order of billions or even trillions. For example, GPT-3 from OpenAI has 175 billion parameters, whereas its more advanced successor, GPT-4, is estimated to have 1.76 trillion parameters.[4] A parameter can be understood as a coefficient learned and tuned during training to minimize the error in predicting the next token for a given sequence of tokens.[5] LLMs are trained on an enormous amount of text to capture the subtleties, intricacies, and grammatical structures of human speech and writing. For example, GPT-3 was trained on a corpus of approximately 500 billion tokens (Brown et al. 2020).

---

[3] For comprehensive review of the technical aspects of LLMs, please see Zhao et al. (2023).
[4] https://the-decoder.com/gpt-4-architecture-datasets-costs-and-more-leaked/

Even though the history of LLMs can be traced back to early developments in neural networks, a real breakthrough was achieved by introducing the transformer architecture (Vaswani et al. 2017). The main innovation of the transformer architecture lies in its self-attention mechanism, which enables the model to weigh the importance of each token in a sequence by considering its interactions with all other tokens in the sequence. This allows the model to understand the context and relationships between tokens in a sequence. The self-attention mechanism is analogous to how we, as human readers, deduce the meaning of an unfamiliar word by looking at its surrounding words to provide context.

Common LLMs based on the transformer architecture include BERT and GPT models. Even though both leverage the transformer architecture, they are designed with different purposes in mind and thus function in different ways. BERT, short for Bidirectional Encoder Representations from Transformers, is designed to learn contextual representations of input sequences by considering both the left and right context. On the other hand, GPT, standing for Generative Pre-training Transformer (GPT), is designed as an autoregressive model for language generation and functions as a sequential language creation model that predicts one word at a time. Its training process follows a one-way (unidirectional) method, where every new word is only influenced by the words that came before it in the text passage. This process, often called causal language modeling, mimics how humans naturally write or speak in a forward-moving flow.

Thanks to its bidirectional approach, BERT excels at tasks that require a deep understanding of context, such as Named Entity Recognition (NER) and Question Answering (QA). However, BERT often requires fine-tuning to improve its performance on domain-specific text. This is because its pre-training while providing the model with a broad understanding of general language patterns, might not capture all the vocabulary, nuances, and unique characteristics of specialized domains such as finance, legal, medical, or scientific texts.

Designed to generate coherent and contextually relevant text, GPT excels at tasks such as text completion, creative writing, and code generation. Certain GPT models (e.g., GPT-3 and its successors) have demonstrated impressive capabilities to perform tasks they are not specifically trained for, with zero-shot and few-shot learning. In simple terms, zero-shot learning enables the model to perform a task it is not trained for without seeing any example of how the task should be done. On the other hand, few-shot learning allows the model to learn a new task from a few examples. LLMs continually undergo remarkable enhancements. Major LLMs released in 2023 and 2024 are summarized in the following table. [6]

---

[5] Technically, the parameters of an LLM consist of weights, biases, and word embeddings. Analogous to the coefficient of a linear function, weights determine the strength of connections between neurons in different layers of the neural network. Biases allow activation functions to be shifted to the left or right and are similar to the constant of a linear function. Unlike those learned by traditional techniques, such as word2vec or GloVe, word embeddings learned by an LLM during the training process are context-dependent, wherein the final representation of each token is informed by the entire input sequence. This allows the model to capture complexities and subtleties of natural language.

*Table 1 Major LLMs from Other Developers Released in 2023 and 2024*

| Release Date | Model Name | Developer | Additional Information |
|---|---|---|---|
| February 2023 | LlaMA 1 | Meta | Open source, available in various sizes (7, 13, 33, and 65 billion parameters); API available |
| May 2023 | PaLM 2 | Google | 340 billion parameters; API available |
| July 2023 | Claude 2 | Anthropic | Large context window (100,000 tokens) and accepting file uploads like PDF as input |
| July 2023 | LlaMA2 | Meta | Open source, various sizes (7, 13, and 70 billion parameters); API available |
| September 2023 | Falcon | TII, UAE | Open source, with 180 billion parameters |
| September 2023 | Mistral 7B | Mistral AI | Open source, 7.3 billion parameters; outperforming Llama 2 (13B) and Llama 1 (34B) on many benchmarks[7] |
| November 2023 | Grok-1 | xAI | Open source conversational AI chatbot, 314 billion parameters; API also available |
| December 2023 | Gemini 1.0 | Google | Available in three sizes (Nano, Pro, and Ultra) |
| February 2024 | Gemma | Google | Open source, lightweight, available in 2B or 7B parameter versions |
| February 2024 | Gemini 1.5 | Google | One million-token context window for Gemini 1.5Pro |
| March 2024 | Claude 3 | Anthropic | Available in three versions: Haiku, Sonnet, and Opus (in ascending order of capability) |
| April 2024 | Mixtral 8x22B | Mistral AI | An open-source Mixture-of-Experts (SMoE) model with 141 billion total parameters, using 39 billion active parameters |
| April 2024 | LlaMA 3 | Meta | Open source, currently two sizes (8B and 70B parameters) |
| June 2024 | Claude Sonnet 3.5 | Anthropic | Significant improvement in performance over the larger Claude 3 Opus |

The performance of an LLM can be measured using the Massive Multitask Language Understanding (MMLU) benchmark (Hendrycks et al. 2021).[8] This benchmark evaluates the knowledge and problem-solving skills of an LLM across 57 different tasks in subjects such as math, history, and law. At the time of writing, Google Gemini Ultra (CoT) claims to outperform GPT-4 (5-shot) with a score of 90.04, which is even higher than the 89.8 score achieved by expert humans.[9] Since GPT models power ChatGPT, we elaborate on models developed and released by OpenAI in the next section.

---

[6] Compiled from multiple sources.
[7] https://mistral.ai/news/announcing-mistral-7b/
[8] For the most current MMLU ranking, see the leaderboard at https://paperswithcode.com/sota/multi-task-language-understanding-on-mmlu.

**2.2 GPT Models from OpenAI**

OpenAI developed a series of LLMs based on its GPT architecture. Table 2 summarizes these models, including their release dates, number of parameters, and context windows. OpenAI introduced GPT-1, its first transformer-based LLM, in June 2018.[10] With 117 million parameters, this model was trained on a large corpus of publicly available text from the Internet and could perform various tasks, such as textual alignment, sentiment analysis, and semantic similarity analysis. As an improved iteration of GPT-1, GPT-2 could generate coherent text sequences and human-like responses to prompts. However, GPT-2 did not perform well at tasks that required more complex reasoning and/or understanding of the context. These limitations led to the development of GPT-3, demonstrating the ability to perform a wide array of language tasks with little to no task-specific training. It can generate text that is contextually rich and often indistinguishable from human-generated writing. Despite these achievements, GPT-3 still exhibits limitations, such as generating factually incorrect information (a phenomenon known as hallucination) and lacking a true understanding of the text it processes.

*Table 2 GPT Models from OpenAI*

| Model Series | Launch Date | Training Data | Parameters | Context Window |
|---|---|---|---|---|
| GPT-1 | June 8, 2018 | Unknown | 117 million | N/A |
| GPT-2 | February 14, 2019 | Unknown | 1.5 billion | 1,024 |
| GPT-3 | June 11, 2020 | Up to Oct 2019 | 175 billion | 2,048 |
| GPT-3.5 | November 30, 2022 | Up to Sep 2021 | 175 billion | 4,096 |
| GPT-4 | March 14, 2023 | Up to Sep 2021 | Estimated ~ 1.76 trl | 8,192 |
| GPT-4 Turbo | November 6, 2023 | Up to Apr 2023 | Estimated ~ 1.76 trl | 128K |
| GPT-4o | May 13, 2024 | Up to Oct 2023 | Unknown | 128K |

The GPT-3.5 series became widely known when ChatGPT was released in November 2022. This news series builds upon its predecessor by offering enhanced coherence for longer passages, improved contextual understanding for nuanced prompts, refined tone and style adoption, reduced hallucinations and misinformation, and improved instruction following. On November 28, 2022, OpenAI unveiled an enhanced iteration of its GPT model, dubbed "text-davinci-003," building upon the previous "text-davinci-002" model. As of November 30, 2022, both models were categorized by OpenAI under the "GPT-3.5" series. Concurrently, on that day, OpenAI launched ChatGPT, an application driven by a model also finetuned for instruction following "text-davinci-002," making that model another member of the GPT-3.5 series.

On March 14, 2023, OpenAI launched GPT-4, which has a much larger context window of 8,192 tokens and is believed to have more than 1.7 trillion parameters. GPT-4 can take text, speech, and image data as input. In addition, GPT-4 can extract text and other data from web pages when a URL is provided in the prompt. It can also search on the Internet if instructed to do so, and this capability allows it to provide more current information beyond its knowledge cut-off date.

---

[9] https://www.newscientist.com/article/2406746-google-says-its-gemini-ai-outperforms-both-gpt-4-and-expert-humans/
[10] Compiled from various sources.

In November 2023, OpenAI introduced GPT-4 Turbo, which has a 128K context window, enabling it to take more than 300 pages of text as a single input. In May 2024, OpenAI launched GPT-4o ("o" for "omni"), which features major advancements in multimodal capabilities. The model seamlessly integrates text, image, and audio processing, allowing natural interactions across different input types. As OpenAI continues to make rapid progress in developing LLMs, even more powerful and capable models are anticipated to be released in the near future.

Each ChatGPT series represents a family of models with varying capabilities that have evolved and improved.[11] These models acquire their capabilities through three crucial steps. The first step involves self-supervised pre-training on a large dataset. The second step involves instruction finetuning, a process that enhances the model's ability to interpret and follow human instructions accurately. The third step involves Reinforcement Learning from Human Feedback (RLHF), a process where feedback provided by human evaluators guides the model in understanding the relative quality of various responses.

In addition to GPT models, OpenAI provides several models for other purposes. Researchers have also started to use some of them. Table 3 provides an overview of these additional models.[12] For example, "text-embedding-ada-002" can generate text embeddings, which are necessary for many downstream tasks, such as classification and information retrieval. For another example, the "whisper-1" model can process audio data.

It is worth mentioning that other developers may also use "GPT" to name their models. Two notable examples are BloombergGPT (S. Wu et al. 2023) and FinGPT (H. Yang, Liu, and Wang 2023). BloombergGPT is a proprietary LLM from Bloomberg with 50 billion parameters trained on a diverse finance dataset. FinGPT is an open-source LLM framework, developed to democratize access to domain-specific models finetuned on finance data.

*Table 3 Other Models Developed by OpenAI*

| Model | Description | How to Access |
|---|---|---|
| DALL·E | A model that can create or modify images in response to text prompts | Labs interface (https://labs.openai.com/) or via API |
| Whisper | A general-purpose speech recognition model that is capable of multilingual speech recognition and translation | Via API with the "whisper-1" model name or open-source version at https://github.com/openai/whisper |

---

[11] For an excellent summary of how GPT models aquire their capabilities, see https://yaofu.notion.site/How-does-GPT-Obtain-its-Ability-Tracing-Emergent-Abilities-of-Language-Models-to-their-Sources-b9a57ac0fcf74f30a1ab9e3e36fa1dc1

[12] More details about these models can be found at https://platform.openai.com/docs/models.

| Embeddings | A set of models that can convert text into numerical representations | Via API with "text-embedding-3-large" (most capable), "text-embedding-3-small", or "text-embedding-ada-002" as the model name |
| Moderation | A model that can detect sensitive or unsafe text involving hate, threatening, self-harm, sexual, or violent content | Via API with "text-moderation-latest" as the model name for the most capable model |

## 2.3 ChatGPT

The term "ChatGPT" commonly refers to the chatbot application created by OpenAI and the underlying models that drive its capabilities. ChatGPT became immensely popular after its release, reaching one million users in five days. In comparison, it took Instagram 2.5 months to acquire the same number of users. Currently, ChatGPT has 180.5 million users.[13] As shown in Figure 1, public interest in ChatGPT is still upward. The Google Trends data also indicates that the general interest in LLMs is not even a fraction of the interest garnered by ChatGPT.

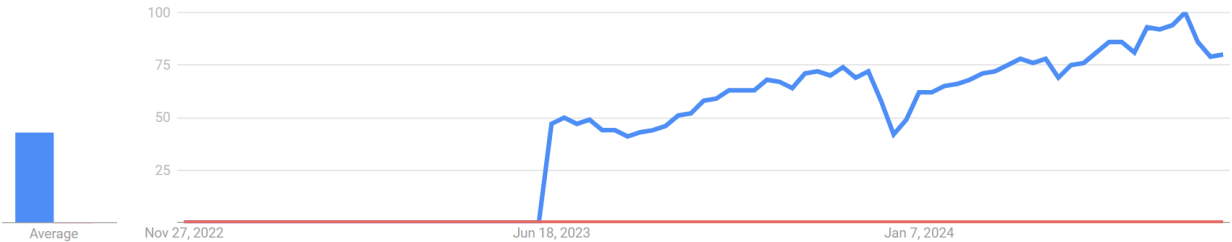

*Figure 1 Google Trend of ChatGPT (Blue) and LLMs (Red)*

The ChatGPT web interface is easy to use and provides an intuitive way to interact with the model behind it. In addition to the free web access to GPT-3.5, OpenAI provides a free playground where users can try various models.[14] GPT-4 is also available via Microsoft's Bing Chat, rebranded as Copilot as of November 2023. However, API is the preferred option for most researchers, especially when processing large datasets. In the appendix, we provide guidance on some of the technical aspects regarding the use of the API.

### III. SCOPE AND METHODOLOGY

In assessing the extant literature, researchers opt for one of two primary review methodologies: a systematic review or a scoping/mapping review (Paré et al. 2015; Yang Li et al. 2018).[15] A systematic review is commonly employed to synthesize established literature on a well-researched topic to elucidate "what works." Conversely, a scoping or mapping review aims to explore the breadth and scope of emerging literature related to a new topic. This review identifies knowledge gaps and informs future research agendas, emphasizing understanding "what has been done" rather than "what has been found."

---

[13] https://explodingtopics.com/blog/chatgpt-users
[14] Available at: https://platform.openai.com/playground
[15] Snyder (2019) classifies literature reviews as systematic, semi-systematic, and integrative. In this paper, we follow the typology of Paré et al. (2015) because it has a more explicit technology-oriented focus.

As indicated by the latest Gartner Hype Cycle (Gartner 2023) and corroborated by anecdotal evidence (Economist 2023b), the adoption of AI/LLMs is apparently in its early stages. Given this context, our study employs a scoping review methodology to survey topics that have been the focus of existing research, identify gaps in the current body of literature, and propose avenues that hold promise for future exploration.

Guided by established protocols in literature review (e.g., Paré et al. 2015; Snyder 2019), we design a procedure that consists of four key steps: 1) formulating a framework to organize and guide the review of the literature, 2) developing and implementing a literature search strategy, which includes eligibility criteria, 3) executing a quality assessment, and 4) interpreting, discussing, and synthesizing the results.

**3.1 Framework and Research Questions**

We develop our framework for organizing and evaluating the studies in our sample by drawing on two streams of research, namely, literature review studies on the adoption of emerging technologies (Yang Li et al. 2018; Lee et al. 2023) and studies on how the state of adoption of an emerging technology influences the type of research that is done or can be done (O'Leary 2008; 2009). Given that ChatGPT is one of the applications of an emerging technology (i.e., Generative AI), we propose a framework that builds on approaches used in recent literature reviews related to the adoption of emerging technologies, such as blockchain and AI (Yang Li et al. 2018; Lee et al. 2023). This means that at a high level, studies can be organized using an input-process-output approach (Lee et al. 2023). Under this approach, *input* relates to motivation for the adoption of new technology or the focus area of application; *process* relates to how the technology is used, what challenges, difficulties, and problems are involved, and what guidelines and best practices are recommended or available; finally, *output* relates to the implications of widespread adoption of the technology.

We complement the input-process-output approach by incorporating evidence from the current stage of LLM adoption. According to O'Leary (2008), the stage of technology adoption determines the type of academic research that can be (is) done. Using the Gartner Hype Cycle (Fenn and Raskino 2008) as a proxy for technology adoption, O'Leary argues that researchers will try to educate themselves about the emerging technology during the early stages when little is known regarding how it works and their capabilities. Over time, as the technology matures and becomes mainstream,[16] we will start seeing large-scale empirical studies focusing on the financial, market, and competitive payoffs from adoption. Combining the insights from these two streams of research, we develop a systematic approach for organizing and reviewing the studies in our sample by focusing on motivation and application focus, process, and output.

---

[16] A technology is considered to have become mainstream when it enters the stage of early majority in the adoption cycle, with an adoption rate of approximately 15% (Stratopoulos, Wang, and Ye 2022).

### 3.1.1 Input: Motivation & Focus Areas

In the technology adoption literature (Rogers 1995), adoption starts with mavericks who can visualize how the new technology could improve the efficiency and effectiveness of their work or their organizations. Extending this logic to accounting and finance research, we argue that researchers who have produced research papers in our sample are those who can visualize the application of LLMs in accounting and finance fields (e.g., financial reporting, audit, and asset pricing). This is consistent with O'Leary (2008), who argues that studies at the early adoption stage will tend to emphasize the positive aspects of the new technology. Therefore, a systematic analysis of the foci of these studies would serve as a proxy for the areas where researchers identify the initial, most accessible opportunities for applying LLMs in accounting and finance.

For this purpose, we first outline the key areas of interest by drawing on the major research areas recognized by the American Accounting Association and American Finance Association, as shown in Table 4. Focusing on motivation and strategic research areas, our framework provides a holistic view of the initial considerations and objectives shaping the use of LLMs in accounting and finance. This also allows us to see the state of the art, gaps, and opportunities for future research in each accounting and finance area.

*Table 4 Accounting & Finance Areas of Research*

| **Accounting** | **Finance** |
| --- | --- |
| Accounting Information Systems | Asset Pricing and Investment |
| Auditing | Corporate Finance |
| Education | Education |
| Financial Accounting and Reporting | Risk Management |
| Management Accounting | |
| Taxation | |

### 3.1.2 Process: How/Capabilities

From the *process* perspective, we examine the specific capabilities of LLMs that researchers leverage in their work. We start by asking ChatGPT 4.0 what capabilities it possesses. At the prompt of "What are you capable of doing?" ChatGPT generates the following list:[17]

- Answering Questions
- Data Analysis and Assistance
- Language Tasks
- Programming and Coding Help
- Creative Content Generation
- Education and Learning

---

[17] Please note that ChatGPT may have customized the answers according to the personal information available in the OpenAI account profile of the author who asked this question.

- General Guidance and Advice

We need to dive deeper into these capabilities to organize the research papers systematically. While ChatGPT's array of functions provides a broad foundation, our study requires a more focused approach tailored to the unique demands of accounting and finance research. For instance, "Language Tasks" is one of the many capabilities of ChatGPT, encompassing translation, grammar checks, writing assistance, and summarization, among others. While these are broadly useful in the context of accounting and finance research (Korinek 2023), some of them are notably more relevant than others, e.g., summarization, which aids in extracting key points from lengthy narratives commonly found in corporate disclosures (A. G. Kim, Muhn, and Nikolaev 2023a). To this end, we propose organizing the capabilities of ChatGPT as follows, each ranked by its complexity and practical application in accounting and finance research:

(1) Word embedding generation.[18]
(2) Information retrieval.
(3) Question-answering (basic) – basic accounting and finance concepts.
(4) Classification and sentiment analysis.[19]
(5) Question-answering (advanced) – application of accounting and finance concepts to complex scenarios.
(6) Text/code generation.
(7) Summarization.
(8) Predictions.
(9) Decision aid by providing recommendations using logical reasoning.

This ranking mirrors the stages and complexity of data analytics, progressing from data management to descriptive, diagnostic, predictive, and prescriptive analytics (Stratopoulos 2018). For example, word embeddings generation and information retrieval represent the initial data collection/management stage, which is essential for subsequent analysis. Classification and sentiment analysis echo diagnostic analytics, as they make it possible to extract deeper insights from data, facilitating a more nuanced understanding of financial narratives. Lastly, capabilities like predictions and logical reasoning are akin to the advanced stages of predictive and prescriptive analytics, where the focus shifts to forecasting future trends and formulating strategic recommendations based on the analyzed data. Understanding the nuances of the process (i.e., capabilities used and their complexity ranking) is important for gaining insights into the practical applications of LLMs in the field and identifying gaps in existing literature and opportunities for future research.

---

[18] A word embedding is a numerical representation of a word as a vector of numbers, such that words closer in the vector space have similar meanings. It is also possible to represent a sentence or even an entire document as a numerical vector, and the results are known as sentence embeddings, document embeddings, or simply text embeddings, which are often more useful for many NLP tasks. We use the term "word embeddings" expansively to include also sentence embeddings, document embeddings, and most broadly text embeddings.

[19] Classification of text involves assigning pre-defined labels to words, sentences, or longer blocks of text. Sentiment analysis also involves a classification task, and the labels are often "positive", "negative", and "neutral", or in a more granular format. We single out sentimental analysis, because we find that many papers use ChatGPT for sentiment analysis.

### 3.1.3 Output: Adoption Maturity & Implications

For the *output* perspective, we analyze the outcomes reported in studies leveraging ChatGPT and similar LLMs in accounting and finance. To distinguish between expected outcomes (where LLMs are used in controlled environments) and actual outcomes (where their intended user base adopts LLMs), we propose the following four groups based on implied adoption maturity (i.e., the stage in the adoption cycle):

(1) Based on general theories and practices, *conceptual papers* represent the researchers' conceptualization and visualization of how the technologies can be applied and how to apply them best.
(2) *Case studies* from early adopters of the technology.
(3) *Potential applications* explore how the technology could be used by demonstrating its capability through large-scale experiments or designing a framework that guides the application of the technology in a specific field.
(4) *Value realization studies* assess the actual impact of technology adoption on organizations and professions. They examine whether integrating the technology has created tangible value for users and organizations.

Mapping these studies into groups of different adoption maturity helps capture the unique aspects of LLMs and their evolving impact over time. Initially, as observed by O'Leary (2008), the adoption of emerging technologies tends to be low, leading to early studies that are conceptual and exploratory, often centered on pilot projects within individual firms. However, with the Economist (2023a) predicting that "Generative AI will go mainstream in 2024," we expect a quicker transition to large-scale empirical studies. These studies will provide insights into the financial and market implications of LLMs, reflecting the evolving nature of this technology.

Another facet of the Output perspective concerns the impact of LLMs on education and the labor market. Several studies have highlighted the potential effects of LLMs on students' future earnings (Huseynov 2023). Coupled with a multitude of studies on the impact of LLMs on white-collar jobs and their broader implications for the labor market, it becomes imperative to review studies that investigate the implications and impacts of LLMs specifically for and on the accounting and finance sectors. This comprehensive approach is necessary to understand the broader implications of LLMs in our domain. By synthesizing these diverse studies, we aim to provide a comprehensive and nuanced view of how LLMs are transforming educational and professional landscapes and redefining the future trajectory of accounting and finance.

### 3.2 Literature Search and Selection

A literature review on an emerging technology aims to create initial conceptualizations and theoretical models rather than examine old ones. Thus, "this type of review often requires a more creative data collection" (Snyder 2019, 336). We endeavor to capture this preliminary conceptualization by reviewing work-in-progress and published research. By including work-in-progress alongside published studies, we can gain insight into the most recent developments and innovative applications of ChatGPT and other LLMs in accounting and finance. This approach enables us to identify the leading edge of research, understand the current state of knowledge, and anticipate the potential transformative effects of these technologies on our field.

We rely on SSRN to identify existing research since most studies are preprints or work-in-progress. SSRN is a popular platform that allows researchers to quickly disseminate their research in social sciences, humanities, and other disciplines. It is a common practice for researchers in accounting, finance, and economics to upload their working papers or recently published papers to SSRN so that the research community can benefit from their most recent or ongoing research findings. This is especially true for studies on emerging and timely topics like LLMs, where rapid advancements necessitate swift knowledge sharing.

To obtain our initial list of working papers, we conducted a search on SSRN for papers that [1] have "ChatGPT" or "GPT" in the title, abstract, or keyword list; [2] are in "Accounting", "Finance", or "Economics" networks; and [3] are uploaded during the period from the beginning of 2022 to the end of March 2024.[20] We exclude working papers with five pages or fewer since such brevity often indicates that the associated studies are likely too preliminary or lack the necessary rigor for meaningful contribution to the field. Among the 264 papers uploaded to SSRN that meet our initial search criteria during our sample period, 37 have an accounting focus, and 46 have a finance focus.

To identify published papers relevant to our research focus, we searched for the World of Science (WoS) platform for papers classified under the "Business Economics" category that contain "ChatGPT" or "GPT" in their titles or abstracts. The "Business Economics" category on WoS encompasses accounting, finance, economics, and other disciplines in business and economics. This search yields an initial list of 144 published papers up to March 31, 2024. Subsequently, we manually reviewed each paper based on its title and abstract, retaining 33 papers relevant to our study, 11 for accounting and 22 for finance.

**3.3 Descriptive Statistics**

Figure 2 depicts the timeline when the working papers were initially uploaded to SSRN. The activity started in late 2022 and has since experienced significant growth, peaking with a notable surge in Spring 2023, followed by some fluctuations after that. The observed pattern from the monthly distribution of uploads is consistent with the unfolding of key developments in LLMs, particularly the release of ChatGPT 3.5 on November 30, 2022, and ChatGPT 4.0 on March 14, 2023.

---

[20] The inclusion of papers from the economics network enables us to identify research that has broad implications for accounting and finance. These papers are addressed in background discussions, for example, when examining the general labor market consequences of LLMs for accounting and finance.

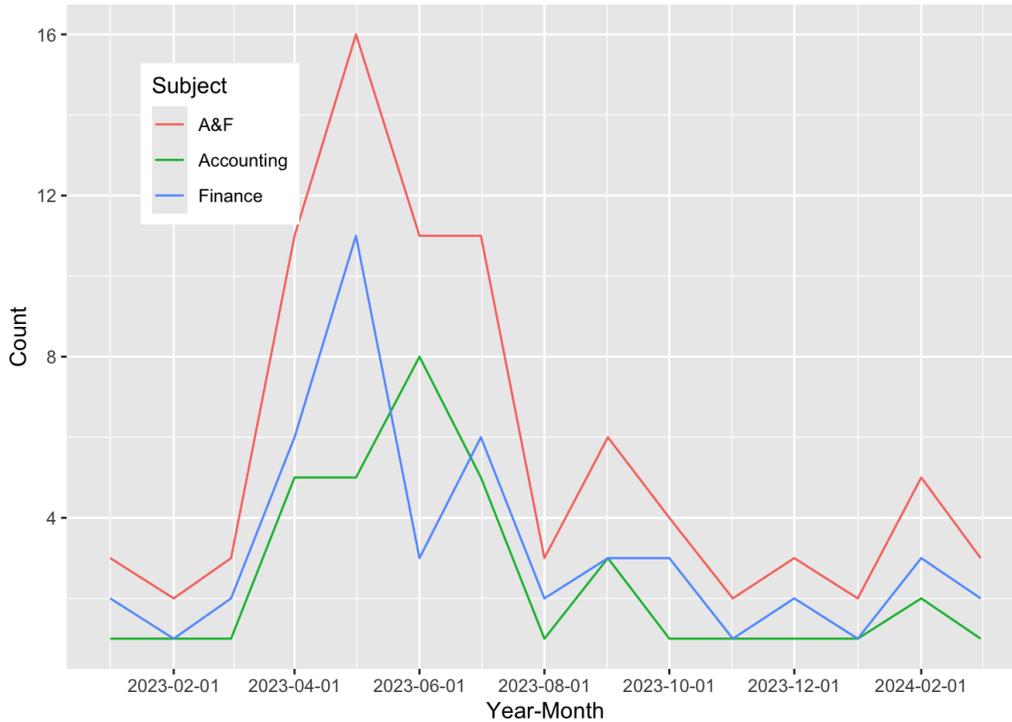

*Figure 2 Uploaded Papers per Month*

The pivotal role of ChatGPT in this body of research is reflected in the frequency of papers using ChatGPT, as shown in Table 5. When a paper discloses the specific model version of ChatGPT used, we further distinguish between GPT-3.5 and GPT-4. In cases where the model version is not disclosed, we use the blanket term "ChatGPT", which, given the timing of these papers, predominantly refers to GPT-3.5. When a paper uses the OpenAI API to access a model, the specific version of the model is often disclosed. Some papers use multiple models, often comparing their performance on certain tasks.

*Table 5 Frequency of Papers by Model Used (Top Ten Models)*

| **Models** | **Count** | **Percent** |
| --- | --- | --- |
| GPT-4 | 26 | 17.7 |
| ChatGPT | 24 | 16.3 |
| GPT-3.5 | 18 | 12.2 |
| GPT-3 | 7 | 4.8 |
| BERT | 5 | 3.4 |
| FinBERT | 5 | 3.4 |
| GPT-3.5-turbo | 5 | 3.4 |
| Google Bard | 4 | 2.7 |
| GPT-2 | 3 | 2.0 |
| Bing Chat | 2 | 1.4 |

Regarding activity within each accounting and finance field (Figure 3), we observe that in the former, most papers focus on financial accounting and reporting (33%) and auditing (18%).

In comparison, in the latter, most papers focus on either Asset Pricing and Investment (51%) or Corporate Finance (22%). From the perspective of our framework, this may indicate that ChatGPT can be more easily applied to these four areas, or the benefit of ChatGPT adoption is more evident in these areas. It could also be the case that more researchers work in financial accounting and reporting than in other areas of accounting research.

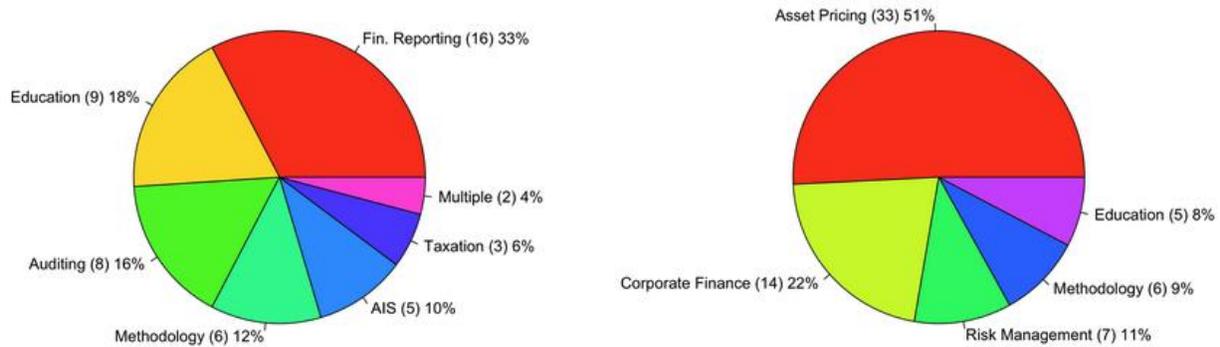

*Figure 3 Most Active Accounting and Finance Fields*

For studies wherein ChatGPT is used as a practical tool, either by researchers or within a corporate setting, we examine the capabilities of ChatGPT being leveraged. Table 6 lists the top five within accounting and finance. It is important to note that these statistics exclude papers in which the authors merely suggest or hypothesize potential uses of ChatGPT or related LLMs without actively or meaningfully employing a model in their research.

*Table 6 Most Frequently Used Tasks in Accounting & Finance*

| Task | Count | Percent |
| --- | --- | --- |
| Sentiment Analysis | 21 | 22.3 |
| Question-answering | 18 | 19.1 |
| Classification | 15 | 16.0 |
| Text generation | 11 | 11.7 |
| Logical reasoning | 10 | 10.6 |

Finally, Tables 7 and 8 capture the frequency of output categories in accounting and finance, respectively. Consistent with the current state of LLM adoption, we see a good representation of conceptual papers. Interestingly, we observe that studies exploring potential applications of ChatGPT and other LLMs in accounting and finance constitute the majority of studies in accounting (57%) and represent over 62% in finance.

*Table 7 Output Categories in Accounting*

| Output Category | Count | Percent |
| --- | --- | --- |
| Potential applications | 38 | 79.2 |
| Conceptual | 9 | 18.8 |
| Case study | 1 | 2.0 |
| ***Total*** | **48** | **100.0** |

Table 8 Output Categories in Finance

| Output Category | Count | Percent |
|---|---|---|
| Potential applications | 44 | 64.7 |
| Conceptual | 13 | 19.1 |
| Value realization | 11 | 16.2 |
| **Total** | **68** | **100.0** |

## IV. SYNTHESIS OF EXTANT LITERATURE

### 4.1 Input Phase: Applications of ChatGPT in Accounting

Practical applications of a new technology start with an awareness and understanding of its functions and capabilities. Researchers can play a significant role in facilitating the widespread adoption of new technologies by demystifying them and envisioning their practical applications. Consistent with the observation of O'Leary (2008), during the early stages, numerous studies aim to educate readers about ChatGPT and LLMs in general. These studies describe and demonstrate their capabilities and explore their potential applications in various accounting fields, such as financial reporting, audit, and tax. For example, Zhao and Wang (2023) discuss the prospective applications of ChatGPT across diverse accounting tasks.[21] Similarly, Street, Wilck, and Chism (2023) encourage CPAs to improve their productivity by experimenting with LLMs for language generation tasks and provide some general principles for safely and effectively doing so. Two studies, Fotoh and Mugwira (2023) and Street and Wilck (2023), provide domain-specific guidelines. The former discusses the potential benefits, pitfalls, and ethical considerations of integrating ChatGPT into external audits, while the latter introduces ChatGPT to forensic accounting professionals.

According to these studies, ChatGPT holds the potential to reshape various accounting processes by automating routine, mundane tasks that require modest professional judgment. Such tasks may include drafting various documents, ranging from internal memos to correspondence with audit or tax clients. LLMs can offer valuable assistance to CPAs working at firms more constrained in resources, helping to address staff shortages exacerbated by the profession's diminishing appeal to new talent (Boritz and Stratopoulos 2023).

However, these potential benefits do not come without risks. The risks or challenges include output accuracy, data security, client privacy, integration complexities, accountability, and intellectual property rights. As such, the authors of these studies urge accounting professionals to maintain a high level of vigilance over these risks. Additionally, the authors caution accounting professionals against an over-reliance on these new technological tools, warning that such over-dependence could potentially undermine their critical thinking skills. Consistent with the findings of Dell'Acqua et al. (2023), they recommend that CPAs exercise professional skepticism and use LLMs to enhance rather than replace their expertise.

---

[21] For existing and proposed applications of ChatGPT in business in general, please see Singh and Singh (2023).

*4.1.1 Auditing*

Audit literature explores the role and effectiveness of audits in enhancing financial statement reliability (audit quality). Major themes include auditor independence, audit risk assessment, internal control evaluation, and ethical challenges in maintaining professional integrity amid regulatory changes, technological advancements, and market pressures (e.g., DeFond and Zhang 2014; Roussy and Perron 2018).

Table 9 presents three clusters of research that explore the use of ChatGPT and LLMs in the auditing domain. The first cluster focuses on applying these technologies to enhance various aspects of the auditing process, including internal, external, and continuous audits. The second cluster examines frameworks and guiding principles for effectively implementing and utilizing these AI technologies in risk-based auditing procedures. The third cluster investigates the cognitive differences between human financial auditors and AI systems and how these differences impact the capabilities and limitations of AI in imitating and assisting human auditors.

*Table 9 Audit Studies*

| Clusters & Studies | Findings |
| --- | --- |
| 1. ***Enhancing Audit Efficiency and Effectiveness with LLMs***: (Emett et al. 2023; Eulerich and Wood 2023; Fotoh and Mugwira 2023; H. Li et al. 2024) | These studies investigate using LLMs to enhance various aspects/processes of internal, external, and continuous auditing. They find that ChatGPT can improve efficiency, automation, and focus on complex tasks, but also raise concerns about accuracy, reliability, and ethical implications. |
| 2. ***Foundations and Frameworks for Integrating LLMs into Audit Processes***: (Föhr, Marten, and Schreyer 2023; Gu et al. 2023; Street and Wilck 2023) | This cluster proposes frameworks and guiding principles for effectively implementing and utilizing deep learning and LLMs, such as ChatGPT, in risk-based auditing procedures. The studies propose holistic frameworks and approaches to lower the auditor's inhibition threshold when applying these technologies. |
| 3. ***Cognitive Differences between Auditors and AI***: (Wei, Wu, and Chu 2023) | This cluster/study explores the heterogeneity in cognitive schemas between human financial auditors and ChatGPT and how these differences in cognitive structures and knowledge organization may impact the competence of AI in imitating and assisting human auditors. |

While these studies provide valuable insights into the potential applications of ChatGPT and LLMs in the auditing domain, it is important to acknowledge certain limitations and gaps that warrant further exploration. The generalizability of these findings across different jurisdictions and audit types remains a pertinent question, necessitating further research to explore the adaptability and scalability of LLMs in auditing practices globally. Moreover, future studies should investigate the impact of ChatGPT and LLMs on audit quality and ethical challenges in maintaining professional integrity in Generative AI-assisted auditing. Overall, the new research aligns with the growing discourse on digital transformation in auditing, challenging traditional methodologies and advocating for reevaluating the auditor's role in the face of technological advancements.

*4.1.2 Financial Accounting and Reporting*

The financial accounting and reporting literature focuses on the development and application of accounting standards, financial statement preparation and presentation, the quality and transparency of financial reporting, the use of accounting information for decision-making (e.g., its impact on capital markets), and the role of ethics and governance in financial reporting and disclosure. Researchers in this area also examine the impacts of technological advancements, sustainability concerns, and globalization on financial accounting and reporting practices (e.g., Barth, Landsman, and Lang 2008; Beyer et al. 2010; Eccles and Krzus 2010; Zeff 2013).

LLMs are well suited for analyzing unstructured textual information, making them valuable for examining various disclosures and reports. As shown in Table 10, LLMs, like ChatGPT, can be leveraged to enhance various aspects of financial reporting, from improving transparency and information quality to aiding in compliance with sustainability disclosures and understanding managerial incentives. The findings suggest that integrating these language models into financial reporting processes can lead to more efficient and informed decision-making by investors and other stakeholders.

*Table 10 - Financial Reporting Studies*

| Clusters & Studies | Findings |
|---|---|
| 1. ***ESG and sustainability Reporting***<br>(Föhr et al. 2023; T. Li, Peng, and Yu 2023; Moreno and Caminero 2023; Ni et al. 2023; Vaghefi et al. 2023; de Villiers, Dimes, and Molinari 2023) | Focusing on the application of Generative AI in ESG and sustainability reporting, these studies find that Generative AI like ChatGPT can facilitate the analysis of such disclosures and improve their quality, highlighting the potential for improved transparency and accountability in corporate sustainability practices. |
| 2. ***Financial Reporting Quality and Transparency***<br>(Andreou, Lambertides, and Magidou 2023; Bai et al. 2023; Bernard et al. 2023; Comlekci et al. 2023; Föhr et al. 2023; Harris 2024) | These studies investigate how LLMs can enhance the quality, transparency, and reliability of financial reporting, and derive valuable insights about companies from their corporate disclosures. LLMs can improve the consistency and clarity of financial disclosures, detect anomalies and risks, and support compliance with accounting standards and regulations. |
| 3. ***Financial Reporting Enrichment and Personalization***<br>(Bai et al. 2023; A. G. Kim, Muhn, and Nikolaev 2023a; Kuroki, Manabe, and Nakagawa 2023) | These studies examine the use of LLMs to analyze financial disclosures, improving investor understanding and decision-making. LLMs can generate detailed, contextualized supplementary information, and relevant and informative financial narratives can enhance user engagement and relevance. |
| 4. ***Corporate Responses to Generative AI & Earnings Forecast***<br>(Jia et al. 2024; E. X. Li, Tu, and Zhou 2023) | Corporations perceive and respond to the emergence of LLMs through adjustments in their strategies, operations, and disclosure practices. These studies highlight managers' heterogeneous views regarding the potential impacts of LLMs and the diverse efforts they are making to adapt to this technological change. |

The emerging literature provides a foundation for understanding the application of LLMs in financial reporting and disclosure, highlighting several practical use cases. Most of these studies focus on employing LLMs to analyze corporate disclosures. However, many opportunities for future research remain. Future studies can explore how LLMs can enhance various aspects of financial analysis, such as extracting and synthesizing relevant information from diverse sources to provide a comprehensive view of a company's financial health and performance or generate valuable insights and recommendations to assist decision-making processes. Furthermore, researchers can investigate the potential of LLMs to streamline transaction recording and financial report generation, potentially automating or semi-automating certain aspects of these processes, such as classifying transactions or generating draft financial reports. This could lead to significant time and cost savings for companies while reducing the risk of human error.

*4.1.3 Taxation*

Taxation literature encompasses tax policy and system design, taxpayer behavior (including compliance, evasion, and tax planning), the impact of taxation on business decisions, and the role of tax incentives. It also examines tax accounting and reporting, the equity and efficiency of tax systems, and the interplay between taxation and economic growth. Researchers have also explored the influence of globalization, digitalization, and sustainability on tax systems and policies (e.g., Hanlon and Heitzman 2010; Devereux and Vella 2014). Despite this broad scope, the application of LLMs in this field has been limited, presenting enormous opportunities for future research.

Choi and Kim (2023) use GPT-4 to analyze tax-audit-related narrative disclosures and construct a novel firm-level measure that captures tax audit periods. They find that tax audits effectively deter firms' tax avoidance, with the effect lingering even after the audits conclude. However, they also document negative firm-level economic consequences of tax audits, such as reduced capital expenditures, increased stock volatility, and decreased debt market access, suggesting that tax audits are an additional source of corporate risk. Two other studies (Alarie et al. 2023; L. Zhang 2023) explicitly focus on tax professionals. These authors test the ability of LLMs to answer tax-related questions and caution tax professionals about the limitations of these models. Alarie et al. (2023) compare the performance of ChatGPT (GPT-3.5 and GPT-4) and Blue J (a proprietary LLM trained on tax data), demonstrating the superiority of Blue J over ChatGPT. One caveat is that the authors of this study are employees of the company that develops and markets Blue J.

*Table 11 - Taxation Papers*

| Clusters & Studies | Findings |
|---|---|
| 1. *Generative AI for Practitioner Tax Research* (Alarie et al. 2023; L. Zhang 2023) | These two studies explore the role of generative AI in practitioner tax research, showing that AI models like ChatGPT can potentially enhance the efficiency of practitioner tax research. However, these models require further improvements before being fully relied on for professional use. |
| 2. *Impact of Tax Audits on Firm Behavior* (Choi and Kim 2023) | The study investigates how firms respond to tax audits based on a measure derived using GPT-4 from tax-audit-related narrative disclosures, showing that audits deter tax avoidance but increase corporate risk and capital constraints. |

Given the small number of studies and the broad scope of taxation literature, there are numerous opportunities for future research. For instance, researchers could examine how LLMs can support tax planning and strategic decision-making for businesses, enabling them to navigate complex tax regulations and optimize their financial strategies. Additionally, future studies could investigate the role of LLMs in detecting and preventing tax evasion, as well as enhancing taxpayer compliance, by leveraging the models' ability to analyze vast amounts of data and identify patterns indicative of fraudulent behavior.

*4.1.4 Accounting Information Systems (AIS)*

AIS literature focuses on the design, implementation, and management of systems that collect, process, and report financial information. Key themes include the integration of technology with accounting processes, the role of AIS in enhancing decision-making and internal controls, data security and privacy issues, and the impact of emerging technologies on accounting practices. It also examines the challenges and benefits of system adoption (Sutton 2010).

Given the predicted exponential growth in enterprise adoption of LLMs (Cooney 2023; Economist 2023a), several studies have examined their role within organizational IT infrastructure and AIS systems. O'Leary (2022; 2023a; 2023b) compared common LLMs and discussed emerging issues, characteristics, risks, and limitations of their enterprise adoption. From a systems development perspective, Beerbaum (2023) and Li and Vasarhelyi (2023) propose integrating LLMs into Robotic Process Automation (RPA) to enhance accounting productivity and efficiency. These technologies can automate routine tasks, allowing accountants to focus on more judgmental work, thus demonstrating the potential for Generative AI-powered tools to transform accounting practices.

*Table 12 AIS Papers*

| Clusters & Studies | Findings |
|---|---|
| 1. *Integrating LLMs in Accounting Processes and Automation* (Beerbaum 2023; H. Li and Vasarhelyi 2023) | These studies investigate the integration of generative AI in accounting systems, finding that AI enhances process automation and efficiency. |
| 2. *Evaluating and Comparing Chatbot Capabilities* (O'Leary 2022; 2023a; 2023b) | These studies explore the commonalities, strengths, and limitations of generative AI-enabled chatbots. The findings emphasize the importance of carefully evaluating the implications of adopting AI chatbots, considering factors such as employee skills, data security, potential misuse, and efficiency gains. |

Despite the valuable insights from these early studies, there are numerous opportunities for further research. Future work could investigate the long-term impact of LLMs on organizational performance, explore the ethical implications of AI-powered decision-making, and examine the necessary organizational changes and employee re-skilling required for effective LLM integration. By addressing these research gaps, researchers can contribute to developing a comprehensive framework for the successful adoption and implementation of LLMs in accounting and management information systems.

## 4.2 Input: Applications of ChatGPT in Finance

Numerous studies argue that LLMs have the potential to transform the way financial professionals do their daily work, given their capabilities of understanding intricate patterns and automating routines and even certain complex processes. Treating ChatGPT as a domain-specific expert is a common method used to visualize the scope and applications of this form of new technology (O'Leary 2008). For example, Zaremba and Demir (2023) explore the applications of ChatGPT in finance and their potential for enhancing NLP-based financial analysis. In multiple related studies, Krause (2023a; 2023b; 2023d) explores the benefits of applying generative AI to finance, ranging from improved operational efficiency to enhanced analytical accuracy. Aldridge (2023) showcases the advantages of AI technologies such as ChatGPT, emphasizing their superior signal extraction capabilities and highlighting their potential for sophisticated financial analysis and forecasting. However, the benefits that ChatGPT can bring to the finance field come with potential risks and other challenges. These may include information inaccuracy, privacy and security concerns, opacity in decision-making processes, labor displacement, and legal considerations (Khan and Umer 2023).

### 4.2.1 Asset Pricing & Investment

Asset pricing and investment literature focuses on the valuation of financial assets, the determinants of asset returns, financial market behavior, and investment strategies. Key themes include developing and testing asset pricing models, the impact of risk factors and market anomalies on asset prices, the role of investor behavior and sentiment in financial decision-making, and the influence of macroeconomic variables on investment performance. The literature also examines portfolio management strategies, market efficiency, and the implications of behavioral finance for asset pricing and investment (e.g., Brunnermeier et al. 2021; Cochrane 2005).

As shown in Table 13, numerous studies highlight how LLMs can be leveraged to enhance return predictability by extracting signals (like sentiment) from corporate disclosures. These models can also support portfolio management, financial advising, and other investment decision-making processes. The findings suggest that integrating LLMs in these areas can lead to more accurate predictions, improved investment outcomes, and data-driven decision-making for investors, portfolio managers, and financial advisors. However, the studies also underscore the importance of addressing potential challenges, such as biases, regulatory uncertainties, and the need for a balanced approach in adopting these technologies.

*Table 13 Asset Pricing & Investment Papers*

| Clusters & Studies | Findings |
|---|---|
| **1. *Impact on market efficiency and asset prices*** (Almeida and Gonçalves 2024; Ante and Demir 2024; Kang, Hwang, and Shin 2024; Nguyen, Nguyen, and Do 2023; Saggu and Ante 2023) | This cluster examines the impact of AI on market efficiency, particularly in crypto markets, and finds that AI innovations, such as ChatGPT, enhance market efficiency and liquidity, suggesting evolving market dynamics. |
| **2. *Investor Sentiment and Predictability*** | This group explores the use of LLMs for sentiment analysis and stock return predictability, |
| (Bond, Klok, and Zhu 2023; Breitung, Kruthof, and Müller 2023; J. Chen et al. 2023; Glasserman and Lin 2023; Kang, Hwang, and Shin 2024; Kirtac and Germano 2024; Lopez-Lira and Tang 2023; Nakano and Yamaoka 2023; Smales 2023; Vamossy and Skog 2023; C. L. Zhang 2023) | revealing that LLMs outperform traditional methods in predicting returns, highlighting the potential for more accurate market insights. |
| **3.** *Portfolio Management and Investment Strategies* (J. Chen et al. 2023; Z. Chen et al. 2023; Fieberg, Hornuf, and Streich 2023; Goyenko and Zhang 2022; Ko and Lee 2024; Lu, Huang, and Li 2023; Romanko, Narayan, and Kwon 2023; Jain et al. 2023; J. H. Kim 2023) | This stream investigates the application of LLMs in portfolio management and investment strategy formulation, finding that LLMs can generate high-return factors and improve portfolio efficiency, suggesting advanced AI capabilities in investment. |
| **4.** *Financial Advisory and Decision Making* (Lo and Ross 2024; Niszczota and Abbas 2023; Oehler and Horn 2024; Pelster and Val 2024) | This cluster examines the role of LLMs in providing financial advice, showing that LLMs can deliver personalized and accurate recommendations, suggesting improved advisory services and decision-making for investors. |
| **5.** *Central Bank Policies and MacroeconomicNews* (Alonso-Robisco and Carbó 2023; Breitung, Kruthof, and Müller 2023; J. Chen et al. 2023; Smales 2023) | This group investigates the use of LLMs in analyzing central bank communications and macroeconomic news, finding that LLMs enhance sentiment measurement and predict market reactions to policy changes. |
| **6.** *AI Applications in Finance and Business* (B. Chen, Wu, and Zhao 2023; Szumilo and Wiegelmann 2024; C. Wang 2023) | This stream explores the broader applications of generative AI in finance and business, highlighting improvements in efficiency, decision-making, and risk management. |

The growing literature on LLMs in asset pricing and investment demonstrates their potential to enhance market efficiency, sentiment analysis, portfolio management, and financial advisory. While these studies provide valuable insights, some limitations should be carefully considered, such as potential AI biases, backtesting challenges, and overreliance on model-generated insights. Future research should address these limitations by developing techniques to identify and mitigate biases in AI models, improving backtesting methodologies for AI-driven strategies, and creating frameworks to balance AI-generated insights with human expertise. Integrating LLMs with traditional financial models could improve predictive power, while longitudinal studies on AI-driven decisions would offer insights into market stability and investor behavior. Exploring LLMs' roles in ESG investing could advance sustainable finance practices.

*4.2.2 Corporate Finance*

Corporate finance literature primarily examines how firms make financial decisions and interact with capital markets. Key themes include capital structure decisions, such as the trade-offs between debt and equity financing, as well as investment and dividend policies. It also examines corporate governance mechanisms and their impact on firm performance, mergers and acquisitions, and the role of financial markets in facilitating corporate growth. Additionally, the literature addresses agency conflict issues and the implications of financial regulations (e.g., Tirole 2010).

Table 14 shows related studies in corporate finance highlight how generative AI, such as ChatGPT, can be leveraged to enhance understanding of corporate policies, firm valuations, and financial industry dynamics. The findings suggest that integrating LLMs in these areas can lead to improved investment, dividend, and employment decisions, as well as better assessment of firm productivity and investor information processing. This could benefit various stakeholders, including corporate managers, investors, and regulators. Additionally, the analysis identifies organizational factors that enable firms to extract greater value from their AI investments, offering insights for successful AI adoption and transformation in corporate settings.

*Table 14 Corporate Finance Papers*

| Clusters & Papers | Findings |
| --- | --- |
| 1. *Corporate Policies* (Álvarez-Díez et al. 2024; Dasgupta, Li, and Wu 2023; Jha et al. 2023; S. Yang 2023) | Studies in this cluster use ChatGPT to extract managerial expectations about corporate policies, such as investment, dividends, and employment. For example, ChatGPT-based investment score provides incremental insights into firms' future investment plans and performance. |
| 2. *Firm Valuation* (Bertomeu et al. 2023; Blomkvist, Qiu, and Zhao 2023; Bughin 2023a; Eisfeldt, Schubert, and Zhang 2023; Gabaix, Koijen, and Yogo 2023; Pietrzak 2023; Wahyono, Rapih, and Boungou 2023) | This cluster examines how LLMs like ChatGPT impact firm valuations, finding greater AI exposure leads to higher returns, suggesting its potential to enhance productivity. The studies also identify key organizational factors that enable "superstar firms" to extract significant value from AI investments, including fit, innovation culture, complementary resources, and competitiveness. |
| 3. *Finance Industry* (Ali and Aysan 2023; Bae et al. 2024; Krause 2023c; Leippold 2023b) | This cluster examines potential financial industry applications of ChatGPT, including loan provisioning, investment analysis, climate finance, and patent valuation. The findings suggest ChatGPT can enhance understanding of corporate policies and enable scalable managerial expectation analysis. |

While the emerging literature provides valuable insights into applying LLMs in corporate finance, certain limitations and gaps remain. Most studies focus on public firms, while private firm analysis using ChatGPT is underexplored. The long-term impacts of widespread AI adoption on firm decisions and market dynamics warrant further study. Researchers could also explore using LLMs to predict and analyze the outcomes of major firm investments, such as mergers and acquisitions. LLMs might provide more accurate forecasts of post-merger performance and integration challenges by processing a wide range of data, including financial statements, cultural indicators, and market sentiment.

### 4.2.3 Risk Management

Risk management literature primarily focuses on identifying, assessing, and mitigating various risks to enhance organizational stability and performance. It explores the impact of behavioral factors on risk perception and decision-making, the integration of risk management into corporate strategy, and the influence of emerging risks, such as cybersecurity threats, climate-related risks, and technological disruptions (e.g., Bromiley et al. 2015; Nocco and Stulz 2006).

Recent studies in Table 15 demonstrate how LLMs, such as ChatGPT, can be leveraged to enhance risk management practices. These findings suggest that integrating LLMs into risk management processes can lead to improved risk measurement, detection of emerging risks, and better-informed decision-making for various stakeholders, including investors, lenders, risk managers, and corporate strategists.

*Table 15 Risk Management Papers*

| Clusters & Papers | Findings |
| --- | --- |
| **1. ChatGPT's Proficiency in Risk Management** (Hofert 2023a; 2023b) | These studies explore ChatGPT's understanding of quantitative risk management and find that it is proficient in non-technical aspects but less so in technical details. |
| **2. *Leveraging Generative AI for Risk Insights*** (A. G. Kim, Muhn, and Nikolaev 2023b; Y. Wang 2023; Z. Wu et al. 2023) | These studies examine the value of LLMs in uncovering corporate risks, finding that AI-generated measures outperform existing ones in predicting stock return volatility and firm responses. |

While these studies demonstrate the promise of integrating LLMs into risk management, some limitations and gaps warrant further research. The findings suggest ChatGPT excels at non-technical risk analysis but struggles with quantitative modeling. Future research could focus on developing hybrid approaches that combine LLMs' natural language processing capabilities with traditional quantitative models. As the scope of the current literature is relatively narrow, exploring applications across diverse risk domains could yield valuable insights. Additionally, as these AI-driven solutions evolve, ensuring their long-term reliability and addressing ethical concerns will require rigorous testing and validation to maintain the integrity and effectiveness of risk management practices.

## 4.3 Process: How ChatGPT is Used in Accounting and Finance Research

A large number of studies employ ChatGPT and related LLMs as research tools. This is not surprising since a significant amount of information provided by or available to accounting/finance professionals comes in the form of unstructured textual data. With their powerful capabilities in NLP, LLMs offer new tools that enable researchers to analyze and draw insights from such textual data. Despite the growing number of studies using ChatGPT as a research tool, ChatGPT and other LLMs are still relatively new to most researchers in the field.

De Kok (2023) proposes a framework for using LLMs such as ChatGPT for textual analysis, focusing on accounting and finance but applicable to other fields. The framework covers model selection, prompt engineering, and validity testing. However, researchers should adapt these guidelines to their specific projects, as LLMs are a rapidly evolving technology that presents ongoing challenges in various research contexts. Several studies explore and propose ways in which LLMs can be used to enhance research productivity for researchers in economics, finance, and other related disciplines (Cao and Zhai 2023; Dowling and Lucey 2023; Feng, Hu, and Li 2023; Korinek 2023). Illustrated using ChatGPT, these studies provide use cases ranging from ideation and literature review to data analysis, coding, and mathematical derivation. The consensus seems that LLMs excel in idea generation and data identification, but they exhibit limitations in literature synthesis and developing suitable testing frameworks. Regarding research communication, Ghio (2024) contends that while AI tools like ChatGPT could challenge English dominance, they may create new power imbalances, reinforcing inequalities rather than democratizing academia.

Next, we discuss the studies that rely on ChatGPT as a research tool by leveraging one or more of its capabilities. While most studies use ChatGPT without fine-tuning, one study in our sample fine-tunes a base model to enhance its capability for domain-specific tasks (B. Zhang, Yang, and Liu 2023). These authors adapted LLaMA-7B using instruction tuning for sentiment analysis of finance text. The fine-tuned model outperforms common LLMs such as FinBERT, ChatGPT, and original LlaMAs in this specific task on finance text. There are other fine-tuned models available for finance text. For a comprehensive discussion of these models and the technical aspects regarding their fine-tuning, see Y. Li et al. (2023).

### *4.3.1 Word Embeddings*

Word embeddings are mathematical representations of words or tokens as real-valued vectors in a high-dimensional space, capturing intricate semantic relationships between them. LLMs use context-dependent word embeddings, wherein the representation of a word depends on its context, thus better capturing the nuanced meanings of and intricate relations between words. As described in the background section, word embeddings are part of the parameters within an LLM and are learned during the training process. Pre-trained LLMs can be used to generate word embeddings.[22]

Recent studies have demonstrated that word embeddings generated by LLMs can outperform traditional methods in various downstream tasks in accounting and finance. Breitung and Müller (2023) introduce a new measure of global business networks using LLM-generated word embeddings, showing their superiority in identifying economic relationships compared to traditional industry classifications. Bandara, Flannery, and Chandak (2023) find that "ada-002" outperforms other models in classifying financial documents, while BERT performs best in predicting earnings surprises using word embeddings from earnings conference calls. Yang (2023) employs "ada-002" to generate context-dependent word embeddings for patent applications, demonstrating their greater power in predicting the economic value of patents. Pungulescu and Stolin (2024) compare BERT, GPT, and semantic fingerprinting to measure the similarity between financial documents and explain stock return correlations.

---

[22] As described in the background section, OpenAI provides three embedding models for this purpose: " text-embedding-3-small", "text-embedding-3-large", and "text-embedding-ada-002". It is important to note that OpenAI is not the sole provider of embedding models. Various other organizations and researchers also offer such models. For those interested in comparing the performance of different embedding models, the Massive Text Embedding Benchmark (MTEB) Leaderboard is an excellent resource. This leaderboard provides a comprehensive ranking of various embedding models based on their performance across multiple tasks and datasets. The MTEB Leaderboard can be accessed at https://huggingface.co/spaces/mteb/leaderboard. It offers valuable insights into the relative strengths of different embedding models, helping researchers and practitioners choose the most suitable model for their specific applications.

As the field continues to evolve, there are numerous opportunities for future research to expand upon these initial findings. Researchers can investigate the performance of word embeddings generated by recent models in a wider range of accounting and finance tasks, such as fraud detection and bankruptcy prediction, to better understand their capabilities and limitations. Future studies can also explore the integration of word embeddings into existing financial analysis methods, potentially developing novel hybrid approaches that combine traditional statistical techniques with the semantic richness captured by LLM-generated embeddings. Furthermore, extending investigations to broader text data, such as annual reports, regulatory filings, and news articles, can uncover new insights. As financial text data grows in volume and complexity, advanced word embedding techniques will become increasingly crucial for extracting meaningful and actionable knowledge.

*4.3.2 Classification*

Classification involves assigning pre-defined labels to the input data based on certain characteristics or criteria. Several recent studies demonstrate the potential of ChatGPT and other LLMs for classifying textual data in accounting and finance research.

One common application is identifying specific topics or types of disclosures in corporate communications. For example, Li, Peng, and Yu (2023) use ChatGPT to identify ESG-related disclosures from conference call transcripts. Kuroki, Manabe, and Nakagawa (2023) use GPT-3.5-turbo to classify management presentations at earnings conference calls of Japanese companies into "facts" or "opinions". Similarly, Jia, Li, Ma, and Xu (2024) employ GPT-4 to determine whether discussions during earnings calls refer to concrete initiatives or general remarks. Moreno and Caminero (2023) use a combination of text mining, ChatGPT, and Google's text-bison to create a Greenhouse Gas Disclosure Index based on companies' ESG reports. Choi and Kim (2023) apply GPT-4 to identify tax audit-related disclosures in annual reports and classify them into different stages. Meng Wang (2023) uses a GPT-based model to identify and classify performance attribution statements made by mutual fund managers in shareholder reports.

Beyond simple disclosure classification, researchers are finding novel applications for using ChatGPT to assess firms' financial characteristics. For instance, Bernard et al. (2023) use GPT-3 to classify XBRL tags and construct a new measure of business complexity. Dasgupta, Li, and Wu (2023) generate a measure of financial flexibility by having ChatGPT evaluate the degree of financial constraint revealed in MD&A sections and the firm's ability to fund investment projects. Notably, Föhr, Marten, and Schreyer (2023) use ChatGPT to classify interview transcripts. This is the first paper we encountered that uses ChatGPT to process qualitative data from field studies. LLMs hold promise as a tool for assisting survey-based research that involves questionnaires (Jansen, Jung, and Salminen 2023).

Overall, these studies highlight the significant potential of ChatGPT and similar LLMs for automating the classification of a wide variety of textual data in accounting and finance. However, researchers should be aware of limitations, such as the potential for inconsistent outputs depending on the prompts used. This is problematic in research contexts where reliability and replicability are crucial. Addressing these challenges is essential to make LLMs more helpful for research applications. As applications expand, these technologies are likely to have a major impact on the field. Future research could explore methods for fine-tuning LLMs on specialized datasets to improve their performance on domain-specific tasks and enhance their understanding of unique terminology and concepts. Developing hybrid models that combine LLMs with other techniques, such as ML models or expert knowledge, may lead to more robust and effective classification systems for accounting and finance applications.

*4.3.3 Sentiment Analysis*

Sentiment analysis is a computational technique used to determine the emotional tone behind a body of text.[23] This type of analysis categorizes a text as having a positive, negative, or neutral tone. Further refinement is possible by assigning numeric values to differentiate the intensity of the tone. Automatic sentiment analysis can be performed using dictionaries or ML approaches. The word list developed by Loughran and McDonald (2011) is a commonly used dictionary for analyzing text in accounting and finance. In recent years, ML-based approaches such as BERT or its fine-tuned version for finance text, i.e., FinBERT (Huang, Wang, and Yang 2023), have gained popularity.

As shown in Table 16, many studies employ ChatGPT and other LLMs for sentiment analysis in various contexts. These studies consistently demonstrate the effectiveness of advanced LLMs like ChatGPT in extracting meaningful sentiment from financial texts, which can be used to forecast market movements or enhance existing financial models. Comparative analyses show that LLMs often outperform traditional methods in financial applications across different languages and markets. While highlighting the significant potential of LLMs to revolutionize financial sentiment analysis and investment decision-making, some studies also address important methodological considerations and challenges, emphasizing the need for careful implementation and domain-specific adaptations in financial applications.

*Table 16 LLMs for Sentiment Analysis*

| Clusters & Papers | Findings |
| --- | --- |
| 1. ***Sentiment Signals for Stock Market Prediction and Trading Strategies*** (Bond, Klok, and Zhu 2023; J. Chen et al. 2023; Lopez-Lira and Tang 2023; Kirtac and Germano 2024; Nakano and Yamaoka 2023; Álvarez-Díez et al. 2024; C. L. Zhang 2023; B. Chen, Wu, and Zhao 2023) | This cluster examines the role of LLMs in sentiment analysis to predict stock market movements and develop trading strategies. The studies demonstrate that LLM-based sentiment analysis can effectively forecast returns and generate profitable trading signals. |
| 2. ***Sentiment Signals for Other Specific Financial Contexts*** (Alonso-Robisco and Carbó 2023; Smales 2023; Bae et al. 2024; Breitung, Kruthof, and Müller | This cluster investigates the application of LLMs for sentiment analysis in various specific financial contexts, including central bank communications, corporate filings, and cryptocurrency markets. The findings demonstrate |

---

[23] Fundamentally, sentiment analysis is a classification task. Because many studies use ChatGPT for sentiment analysis, we group these papers together and discuss them under this dedicated section.

| 2023; Kang, Hwang, and Shin 2024; Andreou, Lambertides, and Magidou 2023) | LLMs' versatility in extracting sentiment-based insights from diverse financial texts. |
|---|---|
| 3. *Comparative Performance of LLMs in Financial Applications* (Hu, Liang, and Yang 2023; Kirtac and Germano 2024; Lopez-Lira and Tang 2023; Bond, Klok, and Zhu 2023; J. Chen et al. 2023) | This cluster compares the performance of various LLMs and traditional methods in financial sentiment analysis. The studies consistently find that advanced LLMs outperform traditional sentiment analysis methods, highlighting their superior ability to capture nuanced sentiments and extract economically relevant information from financial texts. |
| 4. *Methodological Considerations and Challenges* (Leippold 2023a; Glasserman and Lin 2023; Hu, Liang, and Yang 2023) | This cluster addresses methodological issues and challenges in applying LLMs to financial sentiment analysis. The research highlights potential biases and limitations in LLM-based sentiment analysis, emphasizing the need to carefully consider these issues when implementing LLM-based sentiment analysis approaches in finance. |

While many studies employ ChatGPT and other LLMs for sentiment analysis, most focus on sentiment analysis on shorter text, probably due to cost concerns. As LLMs become more affordable, future studies can apply LLMs to sentiment analysis of longer corporate documents, such as full-length 10-Ks and 10-Qs, for a broad sample of firms. Researchers can also apply LLMs to other types of documents, such as social media posts, earnings call transcripts, and analyst reports. While some studies explore the relationship between sentiment and stock returns or other financial variables, more research is still needed on integrating LLMs-enabled sentiment analysis into traditional financial models, such as asset pricing, portfolio optimization, or risk management frameworks.

### *4.3.4 Text Generation*

Text generation involves the automated creation of coherent, contextually appropriate text using advanced algorithms. Generative AI models excel in this task because they are specifically designed for this task. In a practical application, Bai et al. (2023) use LLMs to quantify the extent of new information executives provide during earnings call Q&A sessions, comparing AI-generated answers with executives' responses. They consider executives to have provided little new information if their answers are highly similar to those generated by the LLMs, implicitly assuming that AI cannot provide new information beyond its training data. Wu, Dong, Li, and Shi (2023) use ChatGPT to generate analyses of textual loan assessments, comparing the AI-generated texts with the original human-generated texts. Incorporating text types alongside structured data into deep learning models significantly improves credit default predictions. Harris (2024) employs GPT-4 to expand a list of keywords related to product market competition, generating contextually relevant terms and enabling a more thorough analysis of textual data.

### *4.3.5 Summarization*

Summarization, traditionally a challenging task in NLP, involves condensing extensive textual content into a concise and coherent form while preserving its core information and intent. This process has historically lacked a satisfactory solution due to the complexity of accurately capturing and representing the nuances of large text bodies. However, most advanced LLMs can now create reliable summaries that mirror the depth and even the tone of the original text when instructed to do so. This capability may enable investors and financial analysts to quickly extract useful information from corporate disclosures, promoting market efficiency.

Kim, Muhn, and Nikolaev (2023a) use GPT-3.5-turbo to summarize MD&As in 10-Ks and conference call transcripts. They find that GPT-generated summaries are richer in information content. Bloated disclosures, where originals are excessively long relative to their summaries, can slow price discovery and increase information asymmetry. Similarly, Kim, Muhn, and Nikolaev (2023b) employ GPT-3.5-turbo to summarize and assess a company's exposure to political, climate, and AI-related risks from earnings conference call disclosures. They find that such risk summaries and assessments are informative in that they outperform existing risk measures in predicting stock return volatilities and firms' investment and innovation policies.

*4.3.6 Prediction*

Several studies have focused on the capability of LLMs to make predictions. Li, Tu, and Zhou (2023) find that earnings forecasts generated by GPT-4 exhibit greater forecast errors than analyst consensus, and GPT-4 tends to be more optimistic than financial analysts. Their sub-sample analyses show that GPT-4 generates more accurate forecasts for firms with better information environments or higher-quality disclosures, suggesting its ability to effectively use contextual information. Notably, GPT-4's performance is not affected by text readability, potentially indicating its superior comprehension of complex text. Comlekci et al. (2023) use ChatGPT to forecast future financial performance (i.e., sales and net income) and dividends of public companies included in the Borsa Istanbul 100 Index of Turkey. They find that the performance of ChatGPT significantly improves when recent industry news is provided in addition to historical financial data. This finding highlights the importance of supplying context information to ChatGPT for improved performance.

The implications of these findings are significant for both the financial industry and the development of LLMs. While LLMs demonstrate potential in generating financial forecasts, their performance is not yet on par with human analysts, particularly regarding forecast accuracy. However, the ability of LLMs to effectively utilize contextual information and comprehend complex text suggests that they could serve as valuable tools to augment the work of financial analysts. As LLMs continue to improve and are provided with more comprehensive and diverse data sources, their predictive capabilities may further improve, potentially leading to more accurate and efficient financial forecasting. This could ultimately result in better-informed investment decisions and improved resource allocation in the financial markets.

In summary, evidence from the *process* perspective indicates that newer LLMs outperform traditional methods in many tasks. We note that more studies utilize ChatGPT for tasks of lower complexity (e.g., word embeddings, classification) than for tasks of higher complexity (e.g., summarization and prediction). This discrepancy could be attributed to several factors. First, tasks of lower complexity are often more well-defined and have clearer evaluation metrics, making it easier to assess the performance of LLMs in these areas. Second, tasks of higher complexity may require more specialized knowledge or domain-specific data, which could limit the number of studies exploring these applications. Finally, the relative novelty of advanced LLMs like ChatGPT means that researchers may still be investigating their potential for more complex tasks, and we can expect to see more studies in these areas as the technology matures and gains wider adoption.

### 4.4 Output: Adoption Maturity

Given the current early state of LLM adoption, it is not surprising that numerous conceptual studies aim to educate readers on how to apply LLMs in accounting (Street and Wilck 2023; Street, Wilck, and Chism 2023; J. (Jingwen) Zhao and Wang 2023) and finance settings (e.g., Zaremba and Demir 2023; Romanko, Narayan, and Kwon 2023; Jha et al. 2023). While these papers offer initial and preliminary insights grounded in limited concrete evidence, they are valuable when guidance or best practice is scant at the early stage of a new technology. Drawing on their expert opinions, they visualize the technology's potential scope and applications, critically assessing its strengths and limitations. Such endeavors are essential for promoting a foundational understanding of the technology, covering theoretical underpinnings and practical applications.

With the notable exception of Emett et al. (2023), who offer a comprehensive discussion of the integration of ChatGPT into various internal audit processes within a large multinational company, we have not seen any other case studies. Case studies are essential because they provide real-world insights into how organizations have implemented the technology, including the challenges faced and the lessons learned. These papers are invaluable for understanding the practical aspects of technology adoption and offer a glimpse into the initial stages of its integration into business processes.

Interestingly, there has been a surge in forward-looking studies on how the technology can be utilized, wherein the authors demonstrate its capability through large-scale experiments or develop frameworks that guide the application of the technology to specific fields. We have seen such studies in practically every major field of accounting and finance, e.g., audit (e.g., Gu et al. 2023), financial reporting (e.g., de Villiers, Dimes, and Molinari 2023), asset pricing (e.g., Y. Cheng and Tang 2023), and corporate finance (e.g., Krause 2023c). This group of studies explores potential use cases and envisions how the new technology may create value for users and organizations. In doing so, they serve as a bridge between theoretical understanding and large-scale practical application, highlighting innovative ways technology can be leveraged to improve productivity.

We have seen numerous studies investigating the impact of LLMs on firm valuation. Currently, most of these studies employ market reaction tests to deduce the potential effects on firms or industries based on varying stock price returns observed on key event dates during the development of LLMs. For example, Eisfeldt, Schubert, and Zhang (2023) find that firms with higher exposure to Generative AI experience better returns following ChatGPT's release. Conversely, Blomkvist, Qiu, and Zhao (2023) find negative returns for companies in industries highly susceptible to automation. Bertomeu et al. (2023) investigate the consequences of Italy's temporary ChatGPT ban and observe that Italian firms with greater Generative AI exposure underperform those with less exposure during the ban. Pietrzak (2023) notes minimal market response to companies referencing ChatGPT in their 8-K or 6-K reports, while Bughin (2023b) indicates that only firms with substantial AI investments generate significant returns.

Recent studies find that the education sector has underperformed and received negative investor reactions following the launch of ChatGPT, with traditional education companies being particularly affected (Haugom, Lyocsa, and Halousková 2023; Wahyono, Rapih, and Boungou 2023). This suggests that investors anticipate ChatGPT and similar AI technologies will disrupt the traditional education industry, potentially leading to significant changes in how educational services are delivered and consumed.

Meanwhile, other studies explore the spillover effect of ChatGPT's launch on AI-related crypto assets, observing increased liquidity, efficiency, and returns in AI-Crypto sectors and AI-related crypto assets post-launch (Almeida and Gonçalves 2024; Ante and Demir 2024; Saggu and Ante 2023; Nguyen, Nguyen, and Do 2023). The positive impact on AI-related crypto assets suggests that investors acknowledge the potential of ChatGPT and other advanced AI technologies to drive innovation and growth, leading to increased interest and investment in companies and projects leveraging these technologies.

The studies focusing on market reactions provide a nuanced picture of the impact of LLMs on firm values, highlighting varied effects across sectors and levels of exposure. It is important to note that these studies primarily capture short-term effects. As LLMs become more widely adopted, future research could leverage actual adoption data to assess the return on investment and gain a deeper understanding of their long-term benefits. This presents an opportunity for future studies to explore the long-term impact of LLMs on firm performance, productivity, and competitive advantage. Furthermore, many of the existing studies use broad measures of exposure to LLMs or AI, such as industry-level substitutability or firm-level mentions of ChatGPT from corporate disclosures. As LLMs gain traction and become more integrated into business processes, future research could benefit from utilizing more granular data on actual LLM usage, investment, and integration within firms to better evaluate their impact on performance. This would provide a more accurate and comprehensive understanding of how LLMs contribute to firm value and competitiveness. Additionally, it is worth noting that most current studies focus on U.S. or developed markets. This presents an opportunity for future research to investigate the impact of LLMs on emerging markets and examine potential cross-country differences in adoption and effects. Such research could shed light on the global implications of LLMs and how their impact may vary across different economic, social, and technological contexts.

Our review has revealed a small number of conceptual papers and a substantial body of work focused on exploring potential applications. The substantial volume of the latter work may signify an accelerated adoption of LLMs. Prior literature on technology adoption (e.g., Alexopoulos 2011; Stratopoulos and Wang 2022; Stratopoulos, Wang, and Ye 2022) has introduced several proxies (e.g., book titles, news articles, Google Trend, and firm disclosures) to evaluate the stage of technology adoption. Based on our analysis and evidence related to LLM adoption (Cooney 2023; Economist 2023a), the large number of working papers related to potential applications may be another proxy for predicting the adoption stage.

## 4.5 Output: Impact on Education & Profession

Concerns and debates regarding the impact of technology on the labor market date back to the early 19[th] century, marked by the Luddite movement, and have resurfaced with each new wave of innovation. What sets LLMs apart from prior technologies is their potential to replace highly educated, well-compensated white-collar jobs. This has prompted a debate on job augmentation versus job replacement, and the evidence – at least for now and from the economics literature - is still equivocal (Allen et al. 2023; Hui, Reshef, and Zhou 2023; Kausik 2023; J. Liu et al. 2023). Given that education prepares students for such high-paying jobs, it is not surprising that some college students, especially those from non-STEM majors, feel pessimistic about their future job prospects (Huseynov 2023). Their concerns are consistent with the findings of studies, showing that the stock prices of companies in the education sector reacted negatively to the public release of ChatGPT (Haugom, Lyocsa, and Halousková 2023; Wahyono, Rapih, and Boungou 2023).

Within the realm of accounting and finance, we have observed studies focusing on how LLMs can be used to enhance education or demonstrate the ability of LLMs to master accounting and finance concepts. Liu, Miller and Niu (2023) integrate ChatGPT into a Python programming course for data analytics in finance. Yang and Stivers (2023) assess the ability of ChatGPT and Google Bard to solve undergraduate finance problems, and they find that GPT-4 significantly outperforms Bard-1.0. In an early crowdsourced study, Wood et al. (2023) evaluate the ability of ChatGPT-3.5 to answer accounting questions on various topics and find that ChatGPT overall underperforms accounting students. Bommarito et al. (2023) evaluate the ability of ChatGPT-3.5 and earlier versions to answer CPA exam (multiple-choice) questions on zero-shot prompts. ChatGPT-3.5 significantly underperforms human takers on a sample CPA REG exam, which is heavy in computation. Cheng et al. (2023) find the ability of ChatGPT-3.5 and ChatGPT-4 to solve accounting business cases depends on the type of questions asked. These two models perform better in explaining concepts, applying rules, and evaluating ethical issues than in making journal entries, preparing financial statements, and using software. Abeysekera (2024) finds that ChatGPT-3.5 and ChatGPT-4 demonstrate varied competency levels in solving numerical and narrative-based multiple-choice questions from introductory and advanced financial accounting courses.

A more recent study by Eulerich et al. (2023) tests whether ChatGPT is capable of passing major accounting certification exams, including CPA, CMA (Certified Management Accountant), CIA (Certified Internal Auditor), and EA (Enrolled Agent). They find that while ChatGPT-3.5 cannot pass any of these exams, ChatGPT-4 can pass all of them. They also find that the performance of ChatGPT improves when it is shown some examples or allowed to use a calculator or other resources. These findings are consistent with those from numerous studies that have demonstrated the ability of ChatGPT to perform tasks at a level comparable to a human auditor (e.g., Wei, Wu, and Chu 2023) or augment the abilities of financial analysts (e.g., Gupta 2023).

Evidence from finance studies has shown that ChatGPT can be used for financial advising. For example, Niszczota and Abbas (2023) evaluate the potential of GPT to function as a widely accessible financial robo-advisor. The assessment involves a combination of a financial literacy test and an advice-utilization task known as the Judge-Advisor System. GPT models achieved a satisfactory score. Fieberg et al. (2023) demonstrate that ChatGPT-4 can offer effective financial advice. It can recommend customized investment portfolios tailored to an investor's circumstances, including risk tolerance, risk capacity, and sustainability preferences.

Based on the findings from these studies spanning the supply chain of talents (from education to accounting and investment firms), it is evident that ChatGPT and LLMs, in general, have the potential to disrupt the educational system and the accounting/finance industry. Conceptually, Ballantine, Boyce, and Stoner (2024) argue that the rapid development of AI presents challenges and opportunities for the accounting academy to address its uncritical functionalist view and technical reductionism, necessitating a renewed focus on the human dimension and critical perspectives in accounting education.

## V. DISCUSSION AND RESEARCH OPPORTUNITIES

The main purpose of this paper is to analyze and synthesize the expanding body of working papers and recent publications that focus on the applications and implications of ChatGPT and other LLMs in accounting and finance. Our approach is guided by a forward-looking perspective, aiming to understand the current state of the art thoroughly and identify promising avenues for future research. In our review, we employ a well-structured framework informed by the emerging technology adoption literature to organize and synthesize the vast body of work covering diverse topics. Specifically, we approach the papers from three distinct yet interconnected angles: *input*, which delves into the area of focus and underlying motivations; *process*, exploring the methodologies and capabilities employed; and *output*, assessing the level of adoption maturity and its broader implications.

When synthesizing the literature for each area of the input aspect and most areas of the process aspect, we discussed the limitations and gaps in existing studies and proposed some research opportunities accordingly. In this section, we propose additional research opportunities, most of which are not confined to a specific area but rather span multiple domains or represent overarching themes. We hope these research directions will inspire scholars to pursue novel and impactful studies that advance our understanding of the role and potential of LLMs in accounting and finance. By addressing these research opportunities, the academic community can contribute to developing a more robust and comprehensive body of knowledge, ultimately guiding the responsible and effective adoption of LLMs in practice.

Approaching this body of work from the *input* lens, we have seen studies in practically all accounting and finance areas. While there is a higher concentration in audit, financial reporting, asset valuation, and corporate finance, we note an absence of studies focusing on the potential applications in management accounting/behavioral research. This may be a promising area of future research for several reasons. First, future studies may examine how LLMs can be integrated with Business Intelligence (BI) tools to enhance management accounting by combining complex data analysis with NLP. For example, ChatGPT can interpret outputs from BI systems—like visualizations or statistical data—to generate comprehensive narratives that explain trends and business implications. These narratives may allow decision-makers to understand the context of the numbers. Second, some interest is already in using LLMs in behavioral economics (e.g., Bauer et al. 2023; Tsuchihashi 2023). Researchers may also investigate whether and how reliance on LLMs influences the decision-making process of various parties, such as auditors and equity analysts, from a behavioral perspective. This line of inquiry could explore the potential cognitive biases and heuristics that may arise when professionals interact with LLMs and how these factors might impact the quality and efficiency of their judgments and decisions.

While we have seen studies that propose applications of ChatGPT to the reporting of textual data, we have not seen any studies exploring the application of this new technology to the reporting of financial numbers. This lack of study may be partly attributable to the inadequate capability of ChatGPT in financial reporting, e.g., making journal entries (X. (Joyce) Cheng et al. 2023). As ChatGPT's capability undergoes continuous improvement, there is an opportunity to examine how ChatGPT can be integrated into the accounting information system of a company to automate certain "recording" tasks. In addition to its potential in financial reporting, LLMs could also significantly enhance financial analysis. Researchers could explore the potential of LLMs to enhance financial analysis by providing insights and recommendations based on the analysis of large volumes of financial data. For example, LLMs could be trained on historical financial data and used to identify patterns, trends, and anomalies that may not be easily discernible through traditional analysis methods. This could assist investors in making more informed decisions.

Focusing on the *process* aspect of ChatGPT adoption, we recognize multiple opportunities for future research. In accounting and finance, LLMs can serve as invaluable tools for textual generation. For example, an LLM can effectively define or explain financial concepts like "income statement" or "dividend" in an understandable manner. Such capabilities offer educational advantages and can significantly reduce users' time and effort on manual content creation, providing a cost-effective solution for educational platforms or knowledge bases. On the other hand, the utility of LLMs extends far beyond mere textual generation. They can also provide support for decision-making. LLMs can extract useful information from a large text corpus to assist decision-making for such applications. For instance, an LLM can be trained to analyze earnings announcements, extract key financial metrics, assess management's sentiment, and identify other relevant cues, providing critical input for investment decisions. Unlike text generation applications, decision analytics requires a more nuanced understanding of the context and relies heavily on the model's pattern recognition, inference, and prediction capabilities.

Another fruitful area of future research lies in the use of LLMs as a tool for textual analysis. The evidence from existing studies seems to suggest that ChatGPT has a superior ability as a classifier, which can be used to generate measures for empirical tests. However, most existing studies have applied ChatGPT or other related models to a small volume of textual data, limiting the generalizability and robustness of their findings. Future research should explore the performance of these models on larger and more diverse datasets, spanning different domains, time periods, and document types. Moreover, most studies focus on a single language, predominantly English. As the global economy becomes increasingly more integrated, there is a growing need to extend analysis across diverse languages to gain insights into how regional markets interact and affect the global financial market. We encourage further research focusing on less commonly studied languages to provide a more comprehensive understanding of the multilingual capability of ChatGPT and other LLMs.

Currently, most researchers have used ChatGPT to classify tasks in their studies. ChatGPT and other LLMs are inherently designed to generate text. Text generation thus arguably represents their most significant advantage over more traditional NLP tools. Creating metrics with ChatGPT for empirical testing typically necessitates text classification by assigning categorical labels. However, creating truly novel metrics necessary for addressing intriguing and hitherto unexplored research questions requires a creative application of ChatGPT's capabilities. One way to approach this is to convert a classification task into a text generation task. We have seen two papers that have applied this strategy (A. G. Kim, Muhn, and Nikolaev 2023a; Bai et al. 2023), which holds promise as a path forward for novel research applications using ChatGPT.

Other ChatGPT capabilities have not been explored in existing studies within our review's scope. Two notable examples are Named Entity Recognition (NER) and translation. For example, ChatGPT's capability in NER may be used to refine the disclosure specificity measure proposed by Hope, Hu, and Lu (2016). Using the translation capability, researchers may perform a comparative analysis of the disclosure practices of multinational companies across different geographical areas and jurisdictions. Researchers may also use LLMs to assess the readability of financial text, potentially deriving new metrics that improve on or replace traditional readability measures such as the Fog index and the Bog index.

Current research on ChatGPT has focused primarily on text-based interactions, leaving significant opportunities to explore its capabilities in processing images, audio, and video. As multimodal GAI models like GPT-4o emerge, future studies could evaluate ChatGPT's ability to analyze visual information in financial contexts, extract insights from earnings call recordings, and interpret video content such as CEO interviews. This research could provide valuable insights into the model's ability to integrate information across different modalities and its potential to enhance financial analysis. Expanding research to include multimedia processing capabilities of LLMs like ChatGPT could have significant implications for corporate disclosure practices, investor behavior, regulatory frameworks, and financial technology development. Companies may need to adapt communication strategies to optimize for AI-assisted multimedia content analysis. Understanding how AI-processed multimedia information influences investor decision-making could reshape investment strategies and market dynamics. Insights from multimodal AI analysis could inform new disclosure regulations that account for the impact of non-textual elements in financial communication. Furthermore, research findings could drive the creation of more sophisticated tools for financial analysis that leverage multimodal AI capabilities.

Regarding the *output* aspect of LLM adoption, research in accounting and finance has developed along two main streams: anticipated outcomes and real-world impacts. The first stream includes descriptive or conceptual research proposing potential benefits of LLM adoption, often validated through small-scale experiments with researcher-generated data. This stream also encompasses studies that design frameworks for applying ChatGPT to various accounting and finance fields. The second stream focuses on archival research, examining real-world scenarios where practitioners have integrated LLMs into their operations. These studies aim to identify performance differences between LLM users and non-users, providing practical insights into adoption impacts. This stream also includes research inferring LLM benefits from stock market reactions to major LLM developments.

The second stream focuses on archival research, examining real-world scenarios where practitioners have integrated LLMs into their operations. These studies aim to identify performance differences between LLM users and non-users, providing practical insights into adoption impacts. This stream also includes research inferring LLM benefits from stock market reactions to major LLM developments. As the technology continues to develop and becomes more integrated into accounting and finance practices, researchers can empirically test whether the anticipated benefits of these tools are realized. For example, researchers may investigate the impact of ChatGPT and other LLMs on audit quality, financial reporting quality, or the professional skepticism of practitioners. Before actual data becomes available, researchers may explore the perceived impacts of LLM adoption through surveys among accounting and finance professionals.

Among the papers within our review, there is only one case study, Emett et al. (2023), which discusses how a multinational company has adopted ChatGPT in its internal auditing process. Even though case studies tend to have low external validity, they provide nuanced understanding and contextual insights into real-world applications. We encourage more case studies on generative AI models used in accounting and finance practices. Case studies allow researchers to document and study unique applications of LLMs in different organizations, capturing varied patterns of implementation, challenges faced, and innovative strategies employed. In-depth case analyses reveal how organizational contexts, such as firm size, industry, regulatory environment, and corporate culture, may influence the adoption and impact of LLMs. Case studies may also uncover unintended consequences of technology adoption. In addition, detailed case studies play a pivotal role in theory development and refinement about technology adoption, and insights gained from case studies can also inform future empirical research directions.

Another fruitful area of research is investigating the actual impact of LLMs on firm performance. This can be done by searching corporate disclosures for announcements of LLM adoption and linking the adoption to firm performance, such as stock returns, ROA, or other operational outcomes. LLMs can be a powerful tool to facilitate this type of research. Researchers can use LLMs to sift through vast amounts of textual data in corporate disclosures, press releases, annual reports, and other public communications. Through their advanced text analytics, LLMs can assist in pinpointing which firms have integrated these technologies into their business processes, as well as the context and extent of their adoption. Through such research, academics and practitioners could better understand the strategic value that LLMs bring to firms. Findings from such studies will not only inform corporate decision-making regarding

AI investments but also offer important insights for policymakers, investors, and regulators concerned with the broader economic impacts of LLMs integration into business processes.

In summary, research into the application of ChatGPT and related LLMs within accounting and finance is in its infancy. This burgeoning area of inquiry is abundant with unexplored questions, offering fertile ground for scholarly investigation. To date, the literature has only begun to scratch the surface of these advanced LLMs' potential applications and impacts on the practice of accounting and finance. Given the transformative capabilities of LLMs, there is a clear opportunity for a deeper inquiry into a myriad of pertinent topics. We have proposed a great variety of exciting and promising research avenues, which, if pursued, could yield significant contributions to the theoretical understanding of technology adoption for LLMs as well as their practical applications and implications in various fields of accounting and finance.

## VI. CONCLUDING REMARKS

At a high level, our review of these early studies suggests that integrating LLMs in practically every accounting and finance field can significantly enhance efficiency and effectiveness and that users assisted by LLMs may work more productively than those without such assistance. This points to a plausible trend toward substituting traditional labor with LLM-enhanced workflows. Additionally, the *process* aspect of our review shows that LLMs often outperform traditional methods in tasks like classification, sentiment analysis, and summarization. These superior capabilities of LLMs suggest that researchers and professionals using LLMs may outperform their counterparts relying on older methods. Notably, the *output* aspect of our review indicates a shift in focus from conceptual to practical applications of LLMs. This shift not only demonstrates the growing confidence of researchers and professionals in the capabilities of LLMs but also suggests a potential acceleration in the adoption of LLMs across various domains.

Notwithstanding the pioneering role and contribution of the studies in our review, their primary focus has been proposing ways to perform existing tasks more effectively and efficiently using LLMs like ChatGPT. This pattern is consistent with the early stage of the new technology. However, as highlighted in the seminal work of David (1990), the true transformative potential of a new technology often emerges not merely through the improvement of existing tasks but through a paradigm shift that sees new applications or processes being created. Much like the incremental efficiency gains achieved through the initial use of the dynamo in factories for merely replacing steam engines, over-emphasis on enhancing existing functions with LLMs may yield smaller improvements than these advanced models can. David's analysis suggests that much greater advancements and productivity gains can be achieved when technology is used to reimagine and reengineer processes rather than merely enhance existing ones. Beyond the purview of the studies in our review, the next stride in harnessing LLMs might come from a paradigm shift towards innovative use cases, which entail the creative deployment of LLMs in ways that not only redefine current practices but also unleash new possibilities.

## Appendix Technical Guide

In this appendix, we provide some guidance on how to use ChatGPT as a research tool. These guidelines are based on our ChatGPT experience, reading official OpenAI documentation, posts on the OpenAI Developer Forum, and blog posts from ML practitioners. We also incorporated some good practices we observed from the papers covered in our review.

**Choice of Models**

Section 2.2 provides an overview of the GPT models offered by OpenAI. These models have different capabilities and cost points. You can choose a model that best suits your needs and budget.[24] When you choose a model, you should be aware of the context window to ensure that the length of the input and the intended output fit the context window. Another consideration is whether the performance of the model meets your needs. You can try out different models on the free playground.[25] Once you have chosen a model, it is a good practice to fully disclose the model used, including its series number and other details. For example, if "gpt-3.5-turbo" is specified as the model in an API call, the request will currently be routed to the "gpt-3.5-turbo-0125" model variant behind the scenes. Disclosing the full name of the exact model variant used enables better reproducibility and comparison of research findings over time.

**Context Window**

The context window represents the amount of information an LLM can actively consider when generating a response. It is measured in the number of tokens. For example, the most advanced GPT-4 model has a context window of 128K tokens. However, this does not mean that users can provide a prompt of 128K tokens to the OpenAI API. The context window is shared between the prompt (input) and the response (output). In other words, the 128K-token limit applies to the combined length of the prompt and the generated response. If the prompt alone reaches this limit, the model has no room to generate a response.

It is important to note that tokens are different from words. Tokens include words, punctuation, special characters, line breaks, and even word pieces. As a rule of thumb, 1,000 tokens are roughly equivalent to 750 words. However, the exact number of tokens can vary between models due to their different definitions. OpenAI provides a tool allowing users to find the exact number of tokens for a given input. Users can intuitively see how words differ from tokens on this web interface (https://platform.openai.com/tokenizer). Users can conveniently use the "tiktoken" tokenizer available as a Python library (https://github.com/openai/tiktoken) to count the number of tokens. Knowing the exact number of tokens is important for properly sizing the prompt to avoid exceeding the context window. This information is also useful for estimating the cost based on the pricing scheme. To learn how to count tokens using the "tiktoken" tokenizer, see this OpenAI Jupyter Notebook.[26]

---

[24] The current pricing of various models is available at https://openai.com/pricing. The pricing is based on number of tokens rather than words, and both input tokens and output tokens count as billable tokens.
[25] https://platform.openai.com/playground
[26] https://github.com/openai/openai-cookbook/blob/main/examples/How_to_count_tokens_with_tiktoken.ipynb

Newer LLMs have increasingly larger context windows. However, you may find that the chosen model's context window is insufficient for certain tasks. For example, the most capable GPT-4 model has a context window of 128K, which is equivalent to approximately 100K words. While substantial, this capacity may not be enough to process lengthy documents, such as 10-K filings of some companies, in a single pass for tasks like summarization. A common workaround is to divide a large document into smaller chunks and feed each chunk to the model individually. The outputs from these separate passes can then be aggregated into a combined output, as in Gupta (2023) and Kim, Muhn, and Nikolaev (2023a).

The chunkization approach should work well for classification tasks. However, it is not clear whether this approach works equally well for summarization tasks. Kim, Muhn, and Nikolaev (2023a) use this approach to summarize MD&A and earnings conference call transcripts. They use the summary length relative to that of the original document to capture a construct called "disclosure bloat." It is not clear whether chunkization may introduce bias for long documents. It is plausible that the combined length of the summaries from multiple passes is longer than that of a single-pass summary produced by a model with a sufficiently large context window.

**Major Parameters**

The OpenAI API offers several parameters allowing users to control the output. We encourage researchers to fully disclose their parameter settings when reporting their findings. Such transparency is crucial for enhancing the research's reproducibility and reconciling differences in findings from studies on similar topics.

**Completion length (max_tokens)**: The "max_tokens" parameter allows users to control the output length. The model will stop generating output when the output length reaches the maximum number of tokens set by "max_tokens". Setting a small value for "max_tokens" may result in a truncated output undesired for certain tasks, such as summarization and text generation. For these tasks, a preferred way to control the output length is to expressly tell the model the length limit within the prompt. This way, the model can adjust its output accordingly, ensuring the generated text is complete and coherent.

**Temperature**: This parameter controls the creativity (i.e., randomness) of the output generated by a GPT model. According to the official OpenAI API documentation, the temperature ranges from 0 to 2, with a default value of 1. A higher temperature (e.g., 0.8) allows the model to generate more diverse output, whereas a lower temperature results in more deterministic or focused output. Setting the temperature to zero leads to completely deterministic results. For the papers included in this review, authors often choose a zero temperature to ensure that the output is as deterministic as possible. This choice is consistent with the nature of the research questions being addressed. A zero temperature is recommended for tasks such as classification, summarization, and information extraction to maintain consistency and reproducibility. Empirical evidence supports using a lower temperature for annotation or classification tasks because a lower temperature helps increase consistency without compromising accuracy (Gilardi, Alizadeh, and Kubli 2023).

**TOP_P**: This parameter adjusts the behavior of nucleus sampling, which determines the subset of tokens the model considers when generating the next token. According to the OpenAI API documentation, this parameter ranges from 0 to 2, with a default value 1. A higher value increases the pool of possible subsequent tokens, leading to more creative and unpredictable output. Conversely, a smaller value reduces the pool of possible subsequent tokens, yielding outputs that are more predictable and less diverse. OpenAI generally recommends altering either "temperature" or "top_p" from its default value, but not both simultaneously. This is because both parameters affect the randomness and diversity of the generated output, and adjusting both can lead to unpredictable or undesired results.

**Frequency and Presence Penalty**: These parameters, "frequency_penalty" and "presence_penalty", impose a penalty on the next token depending on how many times or whether the token has already appeared in the output. Both parameters range from -2 to 2 and are defaulted to zero. A negative (positive) value encourages (discourages) repetition. Setting a high value for "frequency_penalty" can help avoid repetition in longer texts. Setting a high value for "presence_penalty" can encourage the introduction of new concepts. For summarization tasks, a slight frequency penalty could be beneficial for reducing redundancy. OpenAI generally recommends altering one, but not both, of these two parameters from their defaults simultaneously. These two parameters are less relevant to classification or information extraction tasks, where the focus is on identifying specific information rather than generating diverse or creative output.

**Logprobs:** The "logprobs" parameter determines whether the model outputs the log probability of each generated token. When set to True, the model will provide the log probability for each token in the output. By default, this parameter is set to False. The log probabilities can be used to assess the model's confidence in its predicted labels for classification tasks. For instance, Bernard et al. (2023) extract the log probabilities from a classification task and use them as inputs for constructing a measure of business complexity. Log probabilities can be used for classification tasks to create precision-recall curves and determine appropriate classification thresholds. For information extraction tasks, the log probabilities can indicate how likely the text contains the extracted information or how likely the model has generated content not present in the original text. By examining the log probabilities, researchers can gauge the reliability of the extracted information and identify potential instances of hallucination or false generation. For more information on how to use log probabilities, see this OpenAI notebook.[27]

**Seed**: The "seed" parameter is a recent addition to the OpenAI API, designed to make the output more deterministic. It accepts an integer and functions similarly to the seed of a random number generator, even though complete determinism is not guaranteed. Using the same seed value and identical parameter settings would likely produce the same output across different requests. To verify or monitor the consistency of the output, users can examine the "system_fingerprint" parameter in the API response. This parameter indicates whether the same seed and parameter settings produce consistent results across multiple requests. We encourage researchers to use this new "seed" parameter for greater reproducibility of their work. For more information on how to use this parameter, see the official guide.[28]

---

[27] https://cookbook.openai.com/examples/using_logprobs

**Look-Ahead Bias**

When using ChatGPT or related LLMs for prediction tasks, it is important to avoid look-ahead bias by being mindful of the knowledge cut-off date of the model used. For example, GPT-4 (currently pointing to "gpt-4-0613") has a knowledge cut-off date of September 2021, which means that GPT-4 was trained on data available up to that date, and it thus has no knowledge of what happened after that. To illustrate this point, Li, Tu, and Zhou (2023) assess the ability of GPT-4 to forecast future earnings of companies and limit their sample of earnings press releases to those announced on or after September 2021. This choice helps mitigate look-ahead bias.

Furthermore, researchers should be aware that OpenAI continuously updates its models and periodically deprecates older iterations, which typically become inaccessible several months later. This means that studies using a specific GPT model for tasks sensitive to the knowledge cut-off date may face challenges when conducting additional analyses during the later review process, as the model used may no longer be available. Researchers should consider this limitation and plan their studies accordingly, ensuring that necessary analyses are completed while the model is still accessible.

**Prompt Engineering**

LLMs take instructions from users in natural human language. Such instructions, known as prompts, play a crucial role in determining the response generated by the model. How an instruction is framed can significantly impact the output from the model. To obtain the desired output, users often need to experiment with different framings of an instruction. This gives rise to a technique known as prompt engineering.

Prompt engineering involves crafting prompts to guide the model's generation of desired outputs. The goal is to enhance output quality for specific tasks by influencing the model's behavior. This technique is crucial for optimizing LLM performance. However, prompt engineering is both an art and a science, and no universal rule fits all contexts. The effectiveness of a prompt depends on various factors, such as the specific task, the model's architecture, and the domain of the input data. Nonetheless, some good practices can help users get started. To learn more about prompt engineering and its best practices, users can refer to the following useful sources:

Best practices for prompt engineering with OpenAI API: https://help.openai.com/en/articles/6654000-best-practices-for-prompt-engineering-with-openai-api

OpenAI guide on prompt engineering: https://platform.openai.com/docs/guides/prompt-engineering

Prompt examples: https://platform.openai.com/examples

---

[28] https://cookbook.openai.com/examples/deterministic_outputs_with_the_seed_parameter

Prompt engineering guide: https://github.com/dair-ai/Prompt-Engineering-Guide

Academic papers on prompt-based tuning for pre-trained LLMs: https://github.com/thunlp/PromptPapers?tab=readme-ov-file#promptpapers

**Other Useful Resources**

OpenAI official documentation: https://platform.openai.com/docs/overview

OpenAI official API reference: https://platform.openai.com/docs/api-reference

OpenAI codebook: https://cookbook.openai.com/

OpenAI Blog Posts: https://openai.com/blog

OpenAI Developer Forum: https://community.openai.com/

**Batch Processing**

Many papers in our review use ChatGPT or other LLMs to process a relatively small amount of data, which will likely reduce cost and processing time. Researchers can employ two strategies to minimize cost and processing time for research questions that require processing a large amount of data. Due to the limitations on the context window and output token size, it is not possible to feed all data to the model at once in many use cases. As a result, a large number of API calls often need to be made, with the prompt repeated for each call. Prompt tokens are charged in the same way as other input tokens. When the prompt is long, it can create a significant overhead since the same prompt needs to be sent multiple times, leading to increased costs and processing time.

To reduce cost and improve efficiency, a recommended approach is to provide a batch of input text for the same prompt. For example, in a sentence classification task, instead of sending one sentence at a time with a prompt, sending a list of sentences (e.g., 100 of them) with the same prompt in a single API call is more cost-effective and time-efficient. The prompt should indicate this to the model and ask the model to return the output in a format easily parsed and matched with each sentence. This batching technique minimizes the required API calls and reduces the overhead associated with repeated prompts.

When dealing with extensive amounts of text data, it is crucial to employ parallel processing techniques rather than sending requests sequentially. OpenAI offers a helpful resource in the form of a notebook demonstrating how to effectively parallelize API requests while implementing throttling mechanisms for managing usage limits. It sends requests concurrently to the API to maximize throughput and reduce processing time. Additionally, it implements throttling techniques to regulate the rate of requests and token usage, ensuring compliance with OpenAI's rate limits. The script includes a retry mechanism that automatically re-attempts failed requests up to a configurable maximum number of times to enhance reliability. The notebook is available at https://github.com/openai/openai-cookbook/blob/main/examples/api_request_parallel_processor.py.

Finally, OpenAI recently introduced a new Batch API that enables users to efficiently process substantial amounts of data through asynchronous batch jobs, offering cost savings and higher rate limits than the standard API. Batch jobs are typically completed within a day, subject to global usage. The Batch API is well-suited for a wide range of tasks involving large datasets, such as classification, sentiment analysis, summarization, and translation.

Users first prepare a batch file in the "JSON" format to utilize the Batch API, where each line corresponds to an individual request. These requests resemble standard Chat Completions API calls, specifying the model, parameters, and messages. The file is then uploaded, and a batch job is created. Once the job is completed, users can retrieve the results. For more information, see https://cookbook.openai.com/examples/batch_processing.

**Declaration of generative AI and AI-assisted technologies in the writing process**

While preparing this manuscript, the author(s) used ChatGPT-4 and Claude 3.5 to improve language and readability. After using these tools, the author(s) reviewed and edited the content as needed and take(s) full responsibility for the publication's content.

# REFERENCES

Abeysekera, Indra. 2024. "ChatGPT And Academia on Accounting Assessments." SSRN Scholarly Paper. Rochester, NY. https://papers.ssrn.com/abstract=4705933.

Alarie, Benjamin, Kim Condon, Susan Massey, and Christopher Yan. 2023. "The Rise of Generative AI for Tax Research." SSRN Scholarly Paper. Rochester, NY. https://papers.ssrn.com/abstract=4476510.

Aldridge, Irene. 2023. "The AI Revolution: From Linear Regression to ChatGPT and Beyond and How It All Connects to Finance." *The Journal of Portfolio Management*, July. https://doi.org/10.3905/jpm.2023.1.519.

Alexopoulos, Michelle. 2011. "Read All about It!! What Happens Following a Technology Shock?" *American Economic Review* 101 (4): 1144–79. https://doi.org/10.1257/aer.101.4.1144.

Ali, Hassnian, and Ahmet Faruk Aysan. 2023. "What Will ChatGPT Revolutionize in Financial Industry?" SSRN Scholarly Paper. Rochester, NY. https://doi.org/10.2139/ssrn.4403372.

Allen, Darcy W. E., Chris Berg, Nataliya Ilyushina, and Jason Potts. 2023. "Large Language Models Reduce Agency Costs." SSRN Scholarly Paper. Rochester, NY. https://doi.org/10.2139/ssrn.4437679.

Almeida, José, and Tiago Cruz Gonçalves. 2024. "The AI Revolution: Are Crypto Markets More Efficient?" SSRN Scholarly Paper. Rochester, NY. https://doi.org/10.2139/ssrn.4757279.

Alonso-Robisco, Andres, and José Manuel Carbó. 2023. "Analysis of CBDC Narrative by Central Banks Using Large Language Models." *Finance Research Letters* 58 (December):104643. https://doi.org/10.1016/j.frl.2023.104643.

Álvarez-Díez, Susana, J. Samuel Baixauli-Soler, Anna Kondratenko, and Gabriel Lozano-Reina. 2024. "Dividend Announcement and the Value of Sentiment Analysis." *Journal of Management Analytics* 0 (0): 1–21. https://doi.org/10.1080/23270012.2024.2306929.

Andreou, Panayiotis C., Neophytos Lambertides, and Marina Magidou. 2023. "Stock Price Crash Risk and the Managerial Rhetoric Mechanism: Evidence from R&D Disclosure in 10-K Filings." SSRN Scholarly Paper. Rochester, NY. https://doi.org/10.2139/ssrn.3891736.

Ante, Lennart, and Ender Demir. 2024. "The ChatGPT Effect on AI-Themed Cryptocurrencies." *Economics and Business Letters* 13 (1): 29–38. https://doi.org/10.17811/ebl.13.1.2024.29-38.

Bae, Junsung, Allen N. Berger, Hyun-Soo Choi, and Hugh Hoikwang Kim. 2024. "Bank Sentiment and Loan Loss Provisioning." SSRN Scholarly Paper. Rochester, NY. https://doi.org/10.2139/ssrn.4745996.

Bai, John (Jianqiu), Nicole M. Boyson, Yi Cao, Miao Liu, and Chi Wan. 2023. "Executives vs. Chatbots: Unmasking Insights through Human-AI Differences in Earnings Conference Q&A." SSRN Scholarly Paper. Rochester, NY. https://doi.org/10.2139/ssrn.4480056.

Ballantine, Joan, Gordon Boyce, and Greg Stoner. 2024. "A Critical Review of AI in Accounting Education: Threat and Opportunity." *Critical Perspectives on Accounting* 99 (March):102711. https://doi.org/10.1016/j.cpa.2024.102711.

Bandara, Wachi, Brandon Flannery, and Anshuma Chandak. 2023. "Can AI Explain Company Performance: A Horserace." SSRN Scholarly Paper. Rochester, NY. https://doi.org/10.2139/ssrn.4480665.

Barth, Mary E., Wayne R. Landsman, and Mark H. Lang. 2008. "International Accounting Standards and Accounting Quality." *Journal of Accounting Research* 46 (3): 467–98. https://doi.org/10.1111/j.1475-679X.2008.00287.x.

Bauer, Kevin, Lena Liebich, Oliver Hinz, and Michael Kosfeld. 2023. "Decoding GPT's Hidden 'Rationality' of Cooperation." SSRN Scholarly Paper. Rochester, NY. https://doi.org/10.2139/ssrn.4576036.

Beerbaum, Dirk Otto. 2023. "Generative Artificial Intelligence (GAI) Ethics Taxonomy- Applying Chat GPT for Robotic Process Automation (GAI-RPA) as Business Case." SSRN Scholarly Paper. Rochester, NY. https://doi.org/10.2139/ssrn.4385025.

Bernard, Darren, Elizabeth Blankespoor, Ties de Kok, and Sara Toynbee. 2023. "Confused Readers: A Modular Measure of Business Complexity." SSRN Scholarly Paper. Rochester, NY. https://doi.org/10.2139/ssrn.4480309.

Bertomeu, Jeremy, Yupeng Lin, Yibin Liu, and Zhenghui Ni. 2023. "Capital Market Consequences of Generative AI: Early Evidence from the Ban of ChatGPT in Italy." SSRN Scholarly Paper. Rochester, NY. https://doi.org/10.2139/ssrn.4452670.

Beyer, Anne, Daniel A. Cohen, Thomas Z. Lys, and Beverly R. Walther. 2010. "The Financial Reporting Environment: Review of the Recent Literature." *Journal of Accounting and Economics* 50 (2): 296–343. https://doi.org/10.1016/j.jacceco.2010.10.003.

Blomkvist, Magnus, Yetaotao Qiu, and Yunfei Zhao. 2023. "Automation and Stock Prices: The Case of ChatGPT." SSRN Scholarly Paper. Rochester, NY. https://doi.org/10.2139/ssrn.4395339.

Bommarito, Jillian, Michael James Bommarito, Jessica Ann Mefford Katz, and Daniel Martin Katz. 2023. "Gpt as Knowledge Worker: A Zero-Shot Evaluation of (AI)CPA Capabilities." SSRN Scholarly Paper. Rochester, NY. https://doi.org/10.2139/ssrn.4322372.

Bond, Shaun A., Hayden Klok, and Min Zhu. 2023. "Large Language Models and Financial Market Sentiment." SSRN Scholarly Paper. Rochester, NY. https://doi.org/10.2139/ssrn.4584928.

Boritz, J. Efrim, and Theophanis C. Stratopoulos. 2023. "AI and the Accounting Profession: Views from Industry and Academia." *Journal of Information Systems* 37 (3): 1–9. https://doi.org/10.2308/ISYS-2023-054.

Breitung, Christian, Garvin Kruthof, and Sebastian Müller. 2023. "Contextualized Sentiment Analysis Using Large Language Models." SSRN Scholarly Paper. Rochester, NY. https://doi.org/10.2139/ssrn.4615038.

Breitung, Christian, and Sebastian Müller. 2023. "Global Business Networks." SSRN Scholarly Paper. Rochester, NY. https://doi.org/10.2139/ssrn.4395079.

Bromiley, Philip, Michael McShane, Anil Nair, and Elzotbek Rustambekov. 2015. "Enterprise Risk Management: Review, Critique, and Research Directions." *Long Range Planning* 48 (4): 265–76. https://doi.org/10.1016/j.lrp.2014.07.005.

Brown, Tom B., Benjamin Mann, Nick Ryder, Melanie Subbiah, Jared Kaplan, Prafulla Dhariwal, Arvind Neelakantan, et al. 2020. "Language Models Are Few-Shot Learners." arXiv. https://doi.org/10.48550/arXiv.2005.14165.

Brunnermeier, Markus, Emmanuel Farhi, Ralph S J Koijen, Arvind Krishnamurthy, Sydney C Ludvigson, Hanno Lustig, Stefan Nagel, and Monika Piazzesi. 2021. "Review Article: Perspectives on the Future of Asset Pricing." *The Review of Financial Studies* 34 (4): 2126–60. https://doi.org/10.1093/rfs/hhaa129.

Bughin, Jacques. 2023a. "Leveraging AI: What Makes Superstar Firms Better Than the Average Corporation?" SSRN Scholarly Paper. Rochester, NY. https://doi.org/10.2139/ssrn.4490396.

———. 2023b. "To ChatGPT or Not to ChatGPT?" SSRN Scholarly Paper. Rochester, NY. https://doi.org/10.2139/ssrn.4411051.
Cao, Yi, and Jia Zhai. 2023. "Bridging the Gap – the Impact of ChatGPT on Financial Research." *Journal of Chinese Economic and Business Studies* 21 (2): 177–91. https://doi.org/10.1080/14765284.2023.2212434.
Chen, Boyang, Zongxiao Wu, and Ruoran Zhao. 2023. "From Fiction to Fact: The Growing Role of Generative AI in Business and Finance." *Journal of Chinese Economic and Business Studies* 21 (4): 471–96. https://doi.org/10.1080/14765284.2023.2245279.
Chen, Jian, Guohao Tang, Guofu Zhou, and Wu Zhu. 2023. "ChatGPT, Stock Market Predictability and Links to the Macroeconomy." SSRN Scholarly Paper. Rochester, NY. https://doi.org/10.2139/ssrn.4660148.
Chen, Zihan, Lei (Nico) Zheng, Cheng Lu, Jialu Yuan, and Di Zhu. 2023. "ChatGPT Informed Graph Neural Network for Stock Movement Prediction." SSRN Scholarly Paper. Rochester, NY. https://doi.org/10.2139/ssrn.4464002.
Cheng, Xu (Joyce), Ryan Dunn, Travis Holt, Kerry Inger, J. Gregory Jenkins, Jefferson Jones, James H. Long, et al. 2023. "Artificial Intelligence's Capabilities, Limitations, and Impact on Accounting Education: Investigating ChatGPT's Performance on Educational Accounting Cases." SSRN Scholarly Paper. Rochester, NY. https://doi.org/10.2139/ssrn.4431202.
Cheng, Yuhan, and Ke Tang. 2023. "GPT's Idea of Stock Factors." SSRN Scholarly Paper. Rochester, NY. https://doi.org/10.2139/ssrn.4560216.
Choi, Ga-Young, and Alex Kim. 2023. "Economic Footprints of Tax Audits: A Generative AI-Driven Approach." SSRN Scholarly Paper. Rochester, NY. https://doi.org/10.2139/ssrn.4645865.
Cochrane, John. 2005. *Asset Pricing: Revised Edition*. Revised edition. Princeton, N.J: Princeton University Press.
Comlekci, İstemi, Serkan Unal, Ali Ozer, and Mehmet Akif Oncu. 2023. "Can AI Technologies Estimate Financials Accurately? A Research on Borsa Istanbul with ChatGPT." SSRN Scholarly Paper. Rochester, NY. https://doi.org/10.2139/ssrn.4545954.
Cooney, Michael. 2023. "Gartner: Top Strategic Technology Trends for 2024." Network World. October 16, 2023. https://www.networkworld.com/article/3708635/gartner-top-strategic-technology-trends-for-2024.html.
Dasgupta, Sudipto, Erica X. N. Li, and Siyuan Wu. 2023. "Inferring Financial Flexibility: Do Actions Speak Louder than Words?" SSRN Scholarly Paper. Rochester, NY. https://doi.org/10.2139/ssrn.4482620.
David, Paul A. 1990. "The Dynamo and the Computer: An Historical Perspective on the Modern Productivity Paradox." *The American Economic Review* 80 (2): 355–61.
DeFond, Mark, and Jieying Zhang. 2014. "A Review of Archival Auditing Research." *Journal of Accounting and Economics*, 2013 Conference Issue, 58 (2): 275–326. https://doi.org/10.1016/j.jacceco.2014.09.002.
Dell'Acqua, Fabrizio, Edward McFowland, Ethan R. Mollick, Hila Lifshitz-Assaf, Katherine Kellogg, Saran Rajendran, Lisa Krayer, François Candelon, and Karim R. Lakhani. 2023. "Navigating the Jagged Technological Frontier: Field Experimental Evidence of the Effects of AI on Knowledge Worker Productivity and Quality." Rochester, NY. https://doi.org/10.2139/ssrn.4573321.

Devereux, Michael P., and John Vella. 2014. "Are We Heading towards a Corporate Tax System Fit for the 21st Century?" *Fiscal Studies* 35 (4): 449–75. https://doi.org/10.1111/j.1475-5890.2014.12038.x.

Dowling, Michael, and Brian Lucey. 2023. "ChatGPT for (Finance) Research: The Bananarama Conjecture." *Finance Research Letters* 53 (May):103662. https://doi.org/10.1016/j.frl.2023.103662.

Eccles, Robert G., and Michael P. Krzus. 2010. *One Report: Integrated Reporting for a Sustainable Strategy*. John Wiley & Sons.

Economist. 2022. "Huge 'Foundation Models' Are Turbo-Charging AI Progress." *The Economist*, June 11, 2022. https://www.economist.com/interactive/briefing/2022/06/11/huge-foundation-models-are-turbo-charging-ai-progress.

———. 2023a. "Generative AI Will Go Mainstream in 2024." *The Economist*, November 13, 2023. https://www.economist.com/the-world-ahead/2023/11/13/generative-ai-will-go-mainstream-in-2024.

———. 2023b. "Your Employer Is (Probably) Unprepared for Artificial Intelligence." *The Economist*, 2023. https://www.economist.com/finance-and-economics/2023/07/16/your-employer-is-probably-unprepared-for-artificial-intelligence.

Eisfeldt, Andrea L., Gregor Schubert, and Miao Ben Zhang. 2023. "Generative AI and Firm Values." SSRN Scholarly Paper. Rochester, NY. https://doi.org/10.2139/ssrn.4436627.

Eloundou, Tyna, Sam Manning, Pamela Mishkin, and Daniel Rock. 2023. "GPTs Are GPTs: An Early Look at the Labor Market Impact Potential of Large Language Models." arXiv.Org. March 17, 2023. https://arxiv.org/abs/2303.10130v4.

Emett, Scott A., Marc Eulerich, Egemen Lipinski, Nicolo Prien, and David A. Wood. 2023. "Leveraging ChatGPT for Enhancing the Internal Audit Process – A Real-World Example from a Large Multinational Company." SSRN Scholarly Paper. Rochester, NY. https://doi.org/10.2139/ssrn.4514238.

Eulerich, Marc, and David A. Wood. 2023. "A Demonstration of How ChatGPT Can Be Used in the Internal Auditing Process." SSRN Scholarly Paper. Rochester, NY. https://doi.org/10.2139/ssrn.4519583.

Feng, Zifeng, Gangqing Hu, and Bingxin Li. 2023. "Unleashing the Power of ChatGPT in Finance Research: Opportunities and Challenges." SSRN Scholarly Paper. Rochester, NY. https://doi.org/10.2139/ssrn.4424979.

Fenn, Jackie, and Mark Raskino. 2008. *Mastering the Hype Cycle: How to Choose the Right Innovation at the Right Time*. Harvard Business Press.

Fieberg, Christian, Lars Hornuf, and David Streich. 2023. "Using GPT-4 for Financial Advice." SSRN Scholarly Paper. Rochester, NY. https://doi.org/10.2139/ssrn.4499485.

Föhr, Tassilo Lars, Kai-Uwe Marten, and Marco Schreyer. 2023. "Deep Learning Meets Risk-Based Auditing: A Holistic Framework for Leveraging Foundation and Task-Specific Models in Audit Procedures." SSRN Scholarly Paper. Rochester, NY. https://doi.org/10.2139/ssrn.4488271.

Föhr, Tassilo Lars, Marco Schreyer, Tatjana Alexandra Juppe, and Kai-Uwe Marten. 2023. "Assuring Sustainable Futures: Auditing Sustainability Reports Using AI Foundation Models." SSRN Scholarly Paper. Rochester, NY. https://doi.org/10.2139/ssrn.4502549.

Fotoh, Lazarus, and Tatenda Mugwira. 2023. "Exploring Large Language Models (ChatGPT) in External Audits: Implications and Ethical Considerations." SSRN Scholarly Paper. Rochester, NY. https://doi.org/10.2139/ssrn.4453835.
Gabaix, Xavier, Ralph S. J. Koijen, and Motohiro Yogo. 2023. "Asset Embeddings." SSRN Scholarly Paper. Rochester, NY. https://doi.org/10.2139/ssrn.4507511.
Gartner. 2023. "What's New in Artificial Intelligence From the 2023 Gartner Hype Cycle." Gartner. August 17, 2023. https://www.gartner.com/en/articles/what-s-new-in-artificial-intelligence-from-the-2023-gartner-hype-cycle.
Ghio, Alessandro. 2024. "Democratizing Academic Research with Artificial Intelligence: The Misleading Case of Language." *Critical Perspectives on Accounting* 98 (January):102687. https://doi.org/10.1016/j.cpa.2023.102687.
Gilardi, Fabrizio, Meysam Alizadeh, and Maël Kubli. 2023. "ChatGPT Outperforms Crowd Workers for Text-Annotation Tasks." *Proceedings of the National Academy of Sciences* 120 (30): e2305016120. https://doi.org/10.1073/pnas.2305016120.
Glasserman, Paul, and Caden Lin. 2023. "Assessing Look-Ahead Bias in Stock Return Predictions Generated By GPT Sentiment Analysis." SSRN Scholarly Paper. Rochester, NY. https://doi.org/10.2139/ssrn.4586726.
Goyenko, Ruslan, and Chengyu Zhang. 2022. "Multi-(Horizon) Factor Investing with AI." SSRN Scholarly Paper. Rochester, NY. https://doi.org/10.2139/ssrn.4187056.
Gu, Hanchi, Marco Schreyer, Kevin Moffitt, and Miklos A. Vasarhelyi. 2023. "Artificial Intelligence Co-Piloted Auditing." SSRN Scholarly Paper. Rochester, NY. https://doi.org/10.2139/ssrn.4444763.
Gupta, Udit. 2023. "GPT-InvestAR: Enhancing Stock Investment Strategies through Annual Report Analysis with Large Language Models." SSRN Scholarly Paper. Rochester, NY. https://doi.org/10.2139/ssrn.4568964.
Hadi, Muhammad Usman, Qasem Al-Tashi, Rizwan Qureshi, Abbas Shah, Amgad Muneer, Muhammad Irfan, Anas Zafar, et al. 2023. *Large Language Models: A Comprehensive Survey of Its Applications, Challenges, Limitations, and Future Prospects*. https://doi.org/10.36227/techrxiv.23589741.v1.
Hanlon, Michelle, and Shane Heitzman. 2010. "A Review of Tax Research." *Journal of Accounting and Economics* 50 (2): 127–78. https://doi.org/10.1016/j.jacceco.2010.09.002.
Harris, Terry. 2024. "Managers' Perception of Product Market Competition and Earnings Management: A Textual Analysis of Firms' 10-K Reports." *Journal of Accounting Literature* ahead-of-print (ahead-of-print). https://doi.org/10.1108/JAL-11-2022-0116.
Haugom, Erik, Stefan Lyocsa, and Martina Halousková. 2023. "The Financial Impact of ChatGPT for the Higher Education Industry in the U.S." SSRN Scholarly Paper. Rochester, NY. https://doi.org/10.2139/ssrn.4573714.
Hendrycks, Dan, Collin Burns, Steven Basart, Andy Zou, Mantas Mazeika, Dawn Song, and Jacob Steinhardt. 2021. "Measuring Massive Multitask Language Understanding." arXiv. https://doi.org/10.48550/arXiv.2009.03300.
Hofert, Marius. 2023a. "Assessing ChatGPT's Proficiency in Quantitative Risk Management." *Risks* 11 (9): 166. https://doi.org/10.3390/risks11090166.
———. 2023b. "Correlation Pitfalls with ChatGPT: Would You Fall for Them?" *Risks* 11 (7): 115. https://doi.org/10.3390/risks11070115.

Hope, Ole-Kristian, Danqi Hu, and Hai Lu. 2016. "The Benefits of Specific Risk-Factor Disclosures." *Review of Accounting Studies* 21 (4): 1005–45. https://doi.org/10.1007/s11142-016-9371-1.
Hu, Nan, Peng Liang, and Xu Yang. 2023. "Whetting All Your Appetites for Financial Tasks with One Meal from GPT? A Comparison of GPT, FinBERT, and Dictionaries in Evaluating Sentiment Analysis." SSRN Scholarly Paper. Rochester, NY. https://papers.ssrn.com/abstract=4426455.
Huang, Allen H., Hui Wang, and Yi Yang. 2023. "FinBERT: A Large Language Model for Extracting Information from Financial Text." *Contemporary Accounting Research* 40 (2): 806–41. https://doi.org/10.1111/1911-3846.12832.
Hui, Xiang, Oren Reshef, and Luofeng Zhou. 2023. "The Short-Term Effects of Generative Artificial Intelligence on Employment: Evidence from an Online Labor Market." SSRN Scholarly Paper. Rochester, NY. https://doi.org/10.2139/ssrn.4544582.
Huseynov, Samir. 2023. "ChatGPT and the Labor Market: Unraveling the Effect of AI Discussions on Students' Earnings Expectations." SSRN Scholarly Paper. Rochester, NY. https://doi.org/10.2139/ssrn.4444728.
Jain, Yash, Shubham Gupta, Serhan Yalciner, Yashodhan Nilesh Joglekar, Parth Khetan, and Tony Zhang. 2023. "Overcoming Complexity in ESG Investing: The Role of Generative AI Integration in Identifying Contextual ESG Factors." SSRN Scholarly Paper. Rochester, NY. https://doi.org/10.2139/ssrn.4495647.
Jansen, Bernard J., Soon-gyo Jung, and Joni Salminen. 2023. "Employing Large Language Models in Survey Research." *Natural Language Processing Journal* 4 (September):100020. https://doi.org/10.1016/j.nlp.2023.100020.
Jha, Manish, Jialin Qian, Michael Weber, and Baozhong Yang. 2023. "ChatGPT and Corporate Policies." SSRN Scholarly Paper. Rochester, NY. https://doi.org/10.2139/ssrn.4521096.
Jia, Ning, Ningzhong Li, Guang Ma, and Da Xu. 2024. "Corporate Responses to Generative AI: Early Evidence from Conference Calls." SSRN Scholarly Paper. Rochester, NY. https://doi.org/10.2139/ssrn.4736295.
Kang, Ryeomyung, Soosung Hwang, and Jinho Shin. 2024. "Cryptocurrency Prices and News Sentiment: Which Exerts Influence on the Other?" SSRN Scholarly Paper. Rochester, NY. https://doi.org/10.2139/ssrn.4714113.
Kausik, B. N. 2023. "Long Tails & the Impact of GPT on Labor." SSRN Scholarly Paper. Rochester, NY. https://doi.org/10.2139/ssrn.4525008.
Khan, Muhammad Salar, and Hamza Umer. 2023. "Chatgpt in Finance: Addressing Ethical Challenges." SSRN Scholarly Paper. Rochester, NY. https://doi.org/10.2139/ssrn.4439967.
Kim, Alex G., Maximilian Muhn, and Valeri V. Nikolaev. 2023a. "Bloated Disclosures: Can ChatGPT Help Investors Process Information?" SSRN Scholarly Paper. Rochester, NY. https://doi.org/10.2139/ssrn.4425527.
———. 2023b. "From Transcripts to Insights: Uncovering Corporate Risks Using Generative AI." SSRN Scholarly Paper. Rochester, NY. https://doi.org/10.2139/ssrn.4593660.
Kim, Jang Ho. 2023. "What If ChatGPT Were a Quant Asset Manager." *Finance Research Letters* 58 (December):104580. https://doi.org/10.1016/j.frl.2023.104580.
Kirtac, Kemal, and Guido Germano. 2024. "Sentiment Trading with Large Language Models." SSRN Scholarly Paper. Rochester, NY. https://doi.org/10.2139/ssrn.4706629.

Ko, Hyungjin, and Jaewook Lee. 2024. "Can ChatGPT Improve Investment Decisions? From a Portfolio Management Perspective." *Finance Research Letters* 64 (June):105433. https://doi.org/10.1016/j.frl.2024.105433.

Kok, Ties de. 2023. "Generative LLMs and Textual Analysis in Accounting: (Chat)GPT as Research Assistant?" SSRN Scholarly Paper. Rochester, NY. https://doi.org/10.2139/ssrn.4429658.

Korinek, Anton. 2023. "Generative AI for Economic Research: Use Cases and Implications for Economists." *Journal of Economic Literature* 61 (4): 1281–1317. https://doi.org/10.1257/jel.20231736.

Krause, David. 2023a. "A Rumsfeldian Framework for Understanding How to Employ Generative AI Models for Financial Analysis." SSRN Scholarly Paper. Rochester, NY. https://doi.org/10.2139/ssrn.4455916.

———. 2023b. "ChatGPT and Generative AI: The New Barbarians at the Gate." SSRN Scholarly Paper. Rochester, NY. https://doi.org/10.2139/ssrn.4447526.

———. 2023c. "ChatGPT and Other AI Models as a Due Diligence Tool: Benefits and Limitations for Private Firm Investment Analysis." SSRN Scholarly Paper. Rochester, NY. https://doi.org/10.2139/ssrn.4416159.

———. 2023d. "Proper Generative AI Prompting for Financial Analysis." SSRN Scholarly Paper. Rochester, NY. https://doi.org/10.2139/ssrn.4453664.

Kuroki, Yutaka, Tomonori Manabe, and Kei Nakagawa. 2023. "Fact or Opinion? – Essential Value for Financial Results Briefing." SSRN Scholarly Paper. Rochester, NY. https://doi.org/10.2139/ssrn.4430511.

Lee, Maggie C. M., Helana Scheepers, Ariel K. H. Lui, and Eric W. T. Ngai. 2023. "The Implementation of Artificial Intelligence in Organizations: A Systematic Literature Review." *Information & Management* 60 (5): 103816. https://doi.org/10.1016/j.im.2023.103816.

Leippold, Markus. 2023a. "Sentiment Spin: Attacking Financial Sentiment with GPT-3." *Finance Research Letters* 55 (July):103957. https://doi.org/10.1016/j.frl.2023.103957.

———. 2023b. "Thus Spoke GPT-3: Interviewing a Large-Language Model on Climate Finance." *Finance Research Letters* 53 (May):103617. https://doi.org/10.1016/j.frl.2022.103617.

Li, Edward Xuejun, Zhiyuan Tu, and Dexin Zhou. 2023. "The Promise and Peril of Generative AI: Evidence from ChatGPT as Sell-Side Analysts." SSRN Scholarly Paper. Rochester, NY. https://doi.org/10.2139/ssrn.4480947.

Li, Huaxia, Marcelo Machado de Freitas, Heejae Lee, and Miklos Vasarhelyi. 2024. "Enhancing Continuous Auditing with Large Language Models: A Framework for Cross-Verification Using Exogenous Textual Data." SSRN Scholarly Paper. Rochester, NY. https://doi.org/10.2139/ssrn.4692960.

Li, Huaxia, and Miklos A. Vasarhelyi. 2023. "Applying Large Language Models in Accounting: A Comparative Analysis of Different Methodologies and Off-the-Shelf Examples." SSRN Scholarly Paper. Rochester, NY. https://doi.org/10.2139/ssrn.4650476.

Li, Tong, Qilin Peng, and Luping Yu. 2023. "ESG Considerations in Acquisitions and Divestitures: Corporate Responses to Mandatory ESG Disclosure." SSRN Scholarly Paper. Rochester, NY. https://doi.org/10.2139/ssrn.4376676.

Li, Yang, Thierry Marier-Bienvenue, Alexis Perron-Brault, Xinyi Wang, and Guy Paré. 2018. "Blockchain Technology in Business Organizations: A Scoping Review." In *Proceedings* *of the 51st Hawaii International Conference on System Sciences*, 4474–83. Hawaii. https://scholarspace.manoa.hawaii.edu/server/api/core/bitstreams/c7c562e3-0a26-4c0a-bc14-e767cf458d7f/content.

Li, Yinheng, Shaofei Wang, Han Ding, and Hang Chen. 2023. "Large Language Models in Finance: A Survey." arXiv. https://doi.org/10.48550/arXiv.2311.10723.

Liu, Jin, Xingchen Xu, Yongjun Li, and Yong Tan. 2023. "'Generate' the Future of Work through AI: Empirical Evidence from Online Labor Markets." SSRN Scholarly Paper. Rochester, NY. https://doi.org/10.2139/ssrn.4529739.

Liu, Yang, Laura K. Miller, and Xu Niu. 2023. "Incorporating ChatGPT into a Financial Data Science Course with Python Programming." SSRN Scholarly Paper. Rochester, NY. https://doi.org/10.2139/ssrn.4412371.

Lo, Andrew W., and Jillian Ross. 2024. "Can ChatGPT Plan Your Retirement?: Generative AI and Financial Advice." SSRN Scholarly Paper. Rochester, NY. https://doi.org/10.2139/ssrn.4722780.

Lopez-Lira, Alejandro, and Yuehua Tang. 2023. "Can ChatGPT Forecast Stock Price Movements? Return Predictability and Large Language Models." *SSRN Electronic Journal*. https://doi.org/10.2139/ssrn.4412788.

Loughran, Tim, and Bill McDonald. 2011. "When Is a Liability Not a Liability? Textual Analysis, Dictionaries, and 10-Ks." *The Journal of Finance* 66 (1): 35–65. https://doi.org/10.1111/j.1540-6261.2010.01625.x.

Lu, Fangzhou, Lei Huang, and Sixuan Li. 2023. "ChatGPT, Generative AI, and Investment Advisory." SSRN Scholarly Paper. Rochester, NY. https://doi.org/10.2139/ssrn.4519182.

Min, Bonan, Hayley Ross, Elior Sulem, Amir Pouran Ben Veyseh, Thien Huu Nguyen, Oscar Sainz, Eneko Agirre, Ilana Heintz, and Dan Roth. 2023. "Recent Advances in Natural Language Processing via Large Pre-Trained Language Models: A Survey." *ACM Computing Surveys* 56 (2): 30:1-30:40. https://doi.org/10.1145/3605943.

Moreno, Angel Ivan, and Teresa Caminero. 2023. "Assessing the Data Challenges of Climate-Related Disclosures in European Banks. A Text Mining Study." SSRN Scholarly Paper. Rochester, NY. https://doi.org/10.2139/ssrn.4592121.

Nakano, Masafumi, and Takuya Yamaoka. 2023. "Enhancing Sentiment Analysis Based Investment by Large Language Models in Japanese Stock Market." SSRN Scholarly Paper. Rochester, NY. https://doi.org/10.2139/ssrn.4511658.

Nguyen, Khanh Quoc, Thanh Huong Nguyen, and Bao Linh Do. 2023. "Narrative Attention and Related Cryptocurrency Returns." *Finance Research Letters* 56 (September):104174. https://doi.org/10.1016/j.frl.2023.104174.

Ni, Jingwei, Julia Bingler, Chiara Colesanti Senni, Mathias Kraus, Glen Gostlow, Tobias Schimanski, Dominik Stammbach, et al. 2023. "chatReport: Democratizing Sustainability Disclosure Analysis through LLM-Based Tools." SSRN Scholarly Paper. Rochester, NY. https://doi.org/10.2139/ssrn.4476733.

Niszczota, Paweł, and Sami Abbas. 2023. "GPT Has Become Financially Literate: Insights from Financial Literacy Tests of GPT and a Preliminary Test of How People Use It as a Source of Advice." *Finance Research Letters* 58 (December):104333. https://doi.org/10.1016/j.frl.2023.104333.

Nocco, Brian W., and René M. Stulz. 2006. "Enterprise Risk Management: Theory and Practice." *Journal of Applied Corporate Finance* 18 (4): 8–20. https://doi.org/10.1111/j.1745-6622.2006.00106.x.

Oehler, Andreas, and Matthias Horn. 2024. "Does ChatGPT Provide Better Advice than Robo-Advisors?" *Finance Research Letters* 60 (February):104898. https://doi.org/10.1016/j.frl.2023.104898.

O'Leary, Daniel E. 2008. "Gartner's Hype Cycle and Information System Research Issues." *International Journal of Accounting Information Systems* 9 (4): 240–52. https://doi.org/10.1016/j.accinf.2008.09.001.

———. 2009. "The Impact of Gartner's Maturity Curve, Adoption Curve, Strategic Technologies on Information Systems Research, with Applications to Artificial Intelligence, ERP, BPM, and RFID." *Journal of Emerging Technologies in Accounting* 6 (January):45–66. https://doi.org/10.2308/jeta.2009.6.1.45.

O'Leary, Daniel E. 2022. "Massive Data Language Models and Conversational Artificial Intelligence: Emerging Issues." *Intelligent Systems in Accounting, Finance and Management* 29 (3): 182–98. https://doi.org/10.1002/isaf.1522.

O'Leary, Daniel E. 2023a. "An Analysis of Three Chatbots: BlenderBot, ChatGPT and LaMDA." *Intelligent Systems in Accounting, Finance and Management* 30 (1): 41–54. https://doi.org/10.1002/isaf.1531.

———. 2023b. "Enterprise Large Language Models: Knowledge Characteristics, Risks, and Organizational Activities." *Intelligent Systems in Accounting, Finance and Management* 30 (3): 113–19. https://doi.org/10.1002/isaf.1541.

Paré, Guy, Marie-Claude Trudel, Mirou Jaana, and Spyros Kitsiou. 2015. "Synthesizing Information Systems Knowledge: A Typology of Literature Reviews." *Information & Management* 52 (2): 183–99. https://doi.org/10.1016/j.im.2014.08.008.

Pelster, Matthias, and Joel Val. 2024. "Can ChatGPT Assist in Picking Stocks?" *Finance Research Letters* 59 (January):104786. https://doi.org/10.1016/j.frl.2023.104786.

Pietrzak, Marcin. 2023. "A Trillion Dollars Race – How Chatgpt Affects Stock Prices." SSRN Scholarly Paper. Rochester, NY. https://doi.org/10.2139/ssrn.4586428.

Pungulescu, Crina, and David Stolin. 2024. "Measuring Document Similarity a Comparative Analysis of NLP Methods in Finance." SSRN Scholarly Paper. Rochester, NY. https://doi.org/10.2139/ssrn.4719865.

Ray, Partha Pratim. 2023. "ChatGPT: A Comprehensive Review on Background, Applications, Key Challenges, Bias, Ethics, Limitations and Future Scope." *Internet of Things and Cyber-Physical Systems* 3 (January):121–54. https://doi.org/10.1016/j.iotcps.2023.04.003.

Rogers, Everett M. 1995. *Diffusion of Innovations*. 4th ed. the Free Press, New York.

Romanko, Oleksandr, Akhilesh Narayan, and Roy Kwon. 2023. "ChatGPT-Based Investment Portfolio Selection." SSRN Scholarly Paper. Rochester, NY. https://doi.org/10.2139/ssrn.4538502.

Roussy, Mélanie, and Alexandre Perron. 2018. "New Perspectives in Internal Audit Research: A Structured Literature Review." *Accounting Perspectives* 17 (3): 345–85. https://doi.org/10.1111/1911-3838.12180.

Saggu, Aman, and Lennart Ante. 2023. "The Influence of ChatGPT on Artificial Intelligence Related Crypto Assets: Evidence from a Synthetic Control Analysis." *Finance Research Letters* 55 (July):103993. https://doi.org/10.1016/j.frl.2023.103993.

Siddik, Abu Bakkar, Yong Li, Arshian Sharif, and Javier Cifuentes-Faura. 2023. "The Role of Artificial Intelligence and Chatgpt in Fintech: Prospects, Challenges, and Research Agendas." SSRN Scholarly Paper. Rochester, NY. https://doi.org/10.2139/ssrn.4439965.

Singh, Harjit, and Avneet Singh. 2023. "ChatGPT: Systematic Review, Applications, and Agenda for Multidisciplinary Research." *Journal of Chinese Economic and Business Studies* 21 (2): 193–212. https://doi.org/10.1080/14765284.2023.2210482.

Smales, Lee A. 2023. "Classification of RBA Monetary Policy Announcements Using ChatGPT." *Finance Research Letters* 58 (December):104514. https://doi.org/10.1016/j.frl.2023.104514.

Snyder, Hannah. 2019. "Literature Review as a Research Methodology: An Overview and Guidelines." *Journal of Business Research* 104:333–39. https://doi.org/10.1016/j.jbusres.2019.07.039.

Stratopoulos, Theophanis C. 2018. "Business Value of Information Technology - A Data Analytics Approach." SSRN Scholarly Paper ID 3186759. Rochester, NY: Social Science Research Network. https://dx.doi.org/10.2139/ssrn.3186759.

Stratopoulos, Theophanis C., and Victor Xiaoqi Wang. 2022. "Estimating the Duration of Competitive Advantage from Emerging Technology Adoption." *International Journal of Accounting Information Systems* 47 (December):100577. https://doi.org/10.1016/j.accinf.2022.100577.

Stratopoulos, Theophanis C., Victor Xiaoqi Wang, and Hua (Jonathan) Ye. 2022. "Use of Corporate Disclosures to Identify the Stage of Blockchain Adoption." *Accounting Horizons* 36 (1): 197–220. https://doi.org/10.2308/HORIZONS-19-101.

Street, Daniel, and Joseph Wilck. 2023. "Let's Have a Chat: Principles for the Effective Application of ChatGPT and Large Language Models in the Practice of Forensic Accounting." SSRN Scholarly Paper. Rochester, NY. https://doi.org/10.2139/ssrn.4351817.

Street, Daniel, Joseph Wilck, and Zachariah Chism. 2023. "Six Principles for the Effective Use of ChatGPT and Other Large Language Models in Accounting." SSRN Scholarly Paper. Rochester, NY. https://papers.ssrn.com/abstract=4551289.

Sutton, Steve G. 2010. "A Research Discipline with No Boundaries: Reflections on 20 Years of Defining AIS Research." *International Journal of Accounting Information Systems* 11 (4): 289–96. https://doi.org/10.1016/j.accinf.2010.09.004.

Szumilo, Nikodem, and Thomas Wiegelmann. 2024. "Real Estate Insights AI: Real Estate's New Roommate – the Good, the Bad and the Algorithmic." *Journal of Property Investment & Finance* ahead-of-print (ahead-of-print). https://doi.org/10.1108/JPIF-01-2024-0001.

Tirole, Jean. 2010. *The Theory of Corporate Finance*. Princeton University Press.

Tsuchihashi, Toshihiro. 2023. "Do Ais Dream of Homo Economicus? Answers from Chatgpt." SSRN Scholarly Paper. Rochester, NY. https://doi.org/10.2139/ssrn.4498882.

Vaghefi, Saeid, Qian Wang, Veruska Muccione, Jingwei Ni, Mathias Kraus, Julia Bingler, Tobias Schimanski, et al. 2023. "ChatClimate: Grounding Conversational AI in Climate Science." SSRN Scholarly Paper. Rochester, NY. https://doi.org/10.2139/ssrn.4414628.

Vamossy, Domonkos F., and Rolf Skog. 2023. "EmTract: Extracting Emotions from Social Media." SSRN Scholarly Paper. Rochester, NY. https://doi.org/10.2139/ssrn.3975884.

Vaswani, Ashish, Noam Shazeer, Niki Parmar, Jakob Uszkoreit, Llion Jones, Aidan N. Gomez, Lukasz Kaiser, and Illia Polosukhin. 2017. "Attention Is All You Need." arXiv. https://doi.org/10.48550/arXiv.1706.03762.

Villiers, Charl de, Ruth Dimes, and Matteo Molinari. 2023. "How Will AI Text Generation and Processing Impact Sustainability Reporting? Critical Analysis, a Conceptual Framework and Avenues for Future Research." *Sustainability Accounting, Management and Policy Journal* ahead-of-print (ahead-of-print). https://doi.org/10.1108/SAMPJ-02-2023-0097.

Wahyono, Budi, Subroto Rapih, and Whelsy Boungou. 2023. "Unleashing the Wordsmith: Analysing the Stock Market Reactions to the Launch of ChatGPT in the US Education Sector." *Finance Research Letters* 58:104576. https://doi.org/10.1016/j.frl.2023.104576.

Wang, Chen. 2023. "Can ChatGPT Personalize Index Funds' Voting Decisions?" SSRN Scholarly Paper. Rochester, NY. https://doi.org/10.2139/ssrn.4413315.

Wang, Meng. 2023. "Heads I Win, Tails It's Chance: Mutual Fund Performance Self-Attribution." SSRN Scholarly Paper. Rochester, NY. https://doi.org/10.2139/ssrn.4553526.

Wang, Yanqing. 2023. "Generative AI in Operational Risk Management: Harnessing the Future of Finance." SSRN Scholarly Paper. Rochester, NY. https://doi.org/10.2139/ssrn.4452504.

Wei, Tian, Han Wu, and Gang Chu. 2023. "Is ChatGPT Competent? Heterogeneity in the Cognitive Schemas of Financial Auditors and Robots." *International Review of Economics & Finance* 88 (November):1389–96. https://doi.org/10.1016/j.iref.2023.07.108.

Wikipedia. 2023. "Uncanny Valley." In *Wikipedia*. https://en.wikipedia.org/w/index.php?title=Uncanny_valley&oldid=1180349546.

Wolfram, Stephen. 2023. "What Is ChatGPT Doing … and Why Does It Work?" Stephen Wolfram Writings. February 14, 2023. https://writings.stephenwolfram.com/2023/02/what-is-chatgpt-doing-and-why-does-it-work/.

Wood, David A., Muskan P. Achhpilia, Mollie T. Adams, Sanaz Aghazadeh, Kazeem Akinyele, Mfon Akpan, Kristian D. Allee, et al. 2023. "The ChatGPT Artificial Intelligence Chatbot: How Well Does It Answer Accounting Assessment Questions?" *Issues in Accounting Education*, April, 1–28. https://doi.org/10.2308/ISSUES-2023-013.

Wu, Shijie, Ozan Irsoy, Steven Lu, Vadim Dabravolski, Mark Dredze, Sebastian Gehrmann, Prabhanjan Kambadur, David Rosenberg, and Gideon Mann. 2023. "BloombergGPT: A Large Language Model for Finance." arXiv. https://doi.org/10.48550/arXiv.2303.17564.

Wu, Zongxiao, Yizhe Dong, Yaoyiran Li, and Baofeng Shi. 2023. "Unleashing the Power of Text for Credit Default Prediction: Comparing Human-Generated and AI-Generated Texts." SSRN Scholarly Paper. Rochester, NY. https://doi.org/10.2139/ssrn.4601317.

Yang, Changyu, and Adam Stivers. 2023. "Investigating AI Languages' Ability to Solve Undergraduate Finance Problems." SSRN Scholarly Paper. Rochester, NY. https://doi.org/10.2139/ssrn.4460814.

Yang, Hongyang, Xiao-Yang Liu, and Christina Dan Wang. 2023. "FinGPT: Open-Source Financial Large Language Models." arXiv. https://doi.org/10.48550/arXiv.2306.06031.

Yang, Stephen. 2023. "Predictive Patentomics: Forecasting Innovation Success and Valuation with ChatGPT." SSRN Scholarly Paper. Rochester, NY. https://doi.org/10.2139/ssrn.4482536.

Zaremba, Adam, and Ender Demir. 2023. "ChatGPT: Unlocking the Future of NLP in Finance." SSRN Scholarly Paper. Rochester, NY. https://doi.org/10.2139/ssrn.4323643.

Zeff, Stephen A. 2013. "The Objectives of Financial Reporting: A Historical Survey and Analysis." *Accounting and Business Research* 43 (4): 262–327. https://doi.org/10.1080/00014788.2013.782237.

Zhang, Boyu, Hongyang Yang, and Xiao-Yang Liu. 2023. "Instruct-FinGPT: Financial Sentiment Analysis by Instruction Tuning of General-Purpose Large Language Models." SSRN Scholarly Paper. Rochester, NY. https://doi.org/10.2139/ssrn.4489831.

Zhang, Christopher L. 2023. "Feel the Market: An Attempt to Identify Additional Factor in the Capital Asset Pricing Model (CAPM) Using Generative Pre-Trained Transformer (GPT) and Bidirectional Encoder Representations from Transformers (BERT)." SSRN Scholarly Paper. Rochester, NY. https://doi.org/10.2139/ssrn.4521946.

Zhang, Libin. 2023. "Four Tax Questions for ChatGPT and Other Language Models." SSRN Scholarly Paper. Rochester, NY. https://papers.ssrn.com/abstract=4458628.

Zhao, Joanna (Jingwen), and Xinruo Wang. 2023. "Unleashing Efficiency and Insights: Exploring the Potential Applications and Challenges of ChatGPT in Accounting." *Journal of Corporate Accounting & Finance*. https://doi.org/10.1002/jcaf.22663.

Zhao, Wayne Xin, Kun Zhou, Junyi Li, Tianyi Tang, Xiaolei Wang, Yupeng Hou, Yingqian Min, et al. 2023. "A Survey of Large Language Models." arXiv. https://doi.org/10.48550/arXiv.2303.18223.